\documentclass[prb,aps,epsf,twocolumn,longbibliography]{revtex4-1}
\usepackage{times}
\usepackage{graphicx}
\usepackage{float}
\usepackage{latexsym,amsmath,amssymb,bm,euscript}
\usepackage{color}
\usepackage{epstopdf}
\usepackage[colorlinks=true,linkcolor=blue,citecolor=blue,urlcolor=blue]{hyperref}
\usepackage{hyperref}

\usepackage[justification=raggedright]{caption}
\usepackage{subcaption}
\usepackage{soul}

\usepackage{physics}
\usepackage{mathtools}

\newcommand{\ncmd}{\newcommand} 
\ncmd{\bs}{\boldsymbol}
\ncmd{\note}[1]{{\color{red}{#1}}}
\ncmd{\mc}{\mathcal}
\ncmd{\nn}{\nonumber}
\ncmd{\trans}[1]{#1^\intercal}
\ncmd{\dl}{\delta}
\ncmd{\Dl}{\Delta}
\ncmd{\sig}{\sigma}

\begin{document}

%%% title
\title{Interaction-driven quantum anomalous Hall insulator in Dirac semimetal}

\author{Hongyu Lu$^1$}
\author{Shouvik Sur$^{2, 3}$}
\email{shouvik.sur@rice.edu}
\author{Shou-Shu Gong$^4$}
\email{shoushu.gong$@$buaa.edu.cn}
\author{D. N. Sheng$^5$}
\email{donna.sheng@csun.edu}
\affiliation{$^1$Department of Physics, University of Hong Kong, Pokfulam Road, Hong Kong SAR, China\\
$^2$Department of Physics and Astronomy, Northwestern University, Evanston, IL 60208, USA\\
$^3$Department of Physics and Astronomy, Rice University, Houston, Texas 77005, USA \\
$^4$Department of Physics and Peng Huanwu Collaborative Center for Research and Education, Beihang University, Beijing 100191, China\\
$^5$Department of Physics and Astronomy, California State University Northridge, Northridge, California 91330, USA
}

%%% abstract
\begin{abstract}
The interaction-driven quantum anomalous Hall (QAH) insulator has been sought for a long time in a  Dirac semimetal with linear band
touching points at the Fermi level. 
By combining  exact diagonalization, density matrix renormalization group, and analytical methods, we study a spinless fermion system on the checkerboard lattice with two fold rotational symmetry, which realizes two Dirac band touching points in the absence of interaction. 
At weak coupling, the Dirac semimetal is stable.
At a finite density-density repulsive interaction, we analyze possible symmetry broken states, and find that an QAH state is stabilized when the interaction strength exceeds the energy scale controlling the separation between the Dirac points.
Through numerical simulations, we verify the existence of the  QAH phase with spontaneous time-reversal symmetry breaking and quantized Chern number $C = 1$.
%%
%Finally, we discuss possible material realization of this system and the implication of our results for understanding the interaction-induced QAH phase in other systems, including twisted bilayer graphene.
\end{abstract}

\maketitle

\section{Introduction}
The integer quantum Hall (IQH) effect is the earliest realization of a topological phase of matter, which has gapped bulk excitations and gapless chiral edge states~\cite{prange1990}.
The topological nature of the IQH state is characterized by an integer Chern number, which manifests through  a quantized Hall conductivity ~\cite{thouless1982}.
In conventional IQH effect time-reversal symmetry (TRS) is broken by the applied external magnetic field. 
Haldane showed that magnetic field is not necessary for realizing an IQH state~\cite{haldane1988}, and TRS may be broken spontaneously \cite{raghu2008}.
Such a new type of IQH state which could be realized in the absence of a magnetic field is called a quantum anomalous Hall (QAH) state, which has prominent potential applications in resistance metrology~\cite{gotz2018} and topological quantum computing~\cite{lian2018}.

Both intrinsic ferromagnetism~\cite{liang2013} or magnetic doping~\cite{yu2010} have been utilized for realizing QAH states in recent experiment at sub-kelvin temperatures~\cite{chang2013, checkelsky2014, chang2015}.
Interaction driven spontaneous TRS breaking provides a distinct route to realizing QAH states in correlated matter.
Such QAH states have been proposed  in correlated two-dimensional semimetals~\cite{raghu2008, sun2009, nandkishore2010}.
In Dirac semimetals (DSMs) on the honeycomb and kagom\'{e} lattices, although mean-field studies propose a QAH state at finite interactions~\cite{raghu2008,weeks2010,wen2010,grushin2013,djuric2014}, numerical calculations only find different charge density wave (CDW) insulating states~\cite{garcia2013,jia2013,daghofer2014,guo2014,motruk2015,capponi2015,scherer2015}. 
Interestingly, the QAH state is also predicted to be the dominant instability of spinless  semimetals with a quadratic band touching (QBT) point at the Fermi level~\cite{sun2009, nandkishore2010}. 
Because of the finite density of states at the Fermi level, nearest-neighbor repulsive interactions are marginally relevant and can stabilize a QAH state at weak coupling in such spinless fermion system~\cite{chong2008,sun2008,nandkishore2010,sun2009,wen2010,tsai2015}.
Recently, this QAH state has also been identified at strong coupling by unbiased numerical calculation~\cite{wu2016, zhu2016, sur2018, zeng2018}.
In material simulation, such a QAH state is predicted to exist in the hematite nanosheets, which may have a very large gap $\sim 300$meV~\cite{liang2017}. 
%%
%%

%%%%%%%%%%%%%%%%%%%%%%%%%%%%%%%%%%%%
\begin{figure}[!t]
\centering
\includegraphics[width = 1\columnwidth]{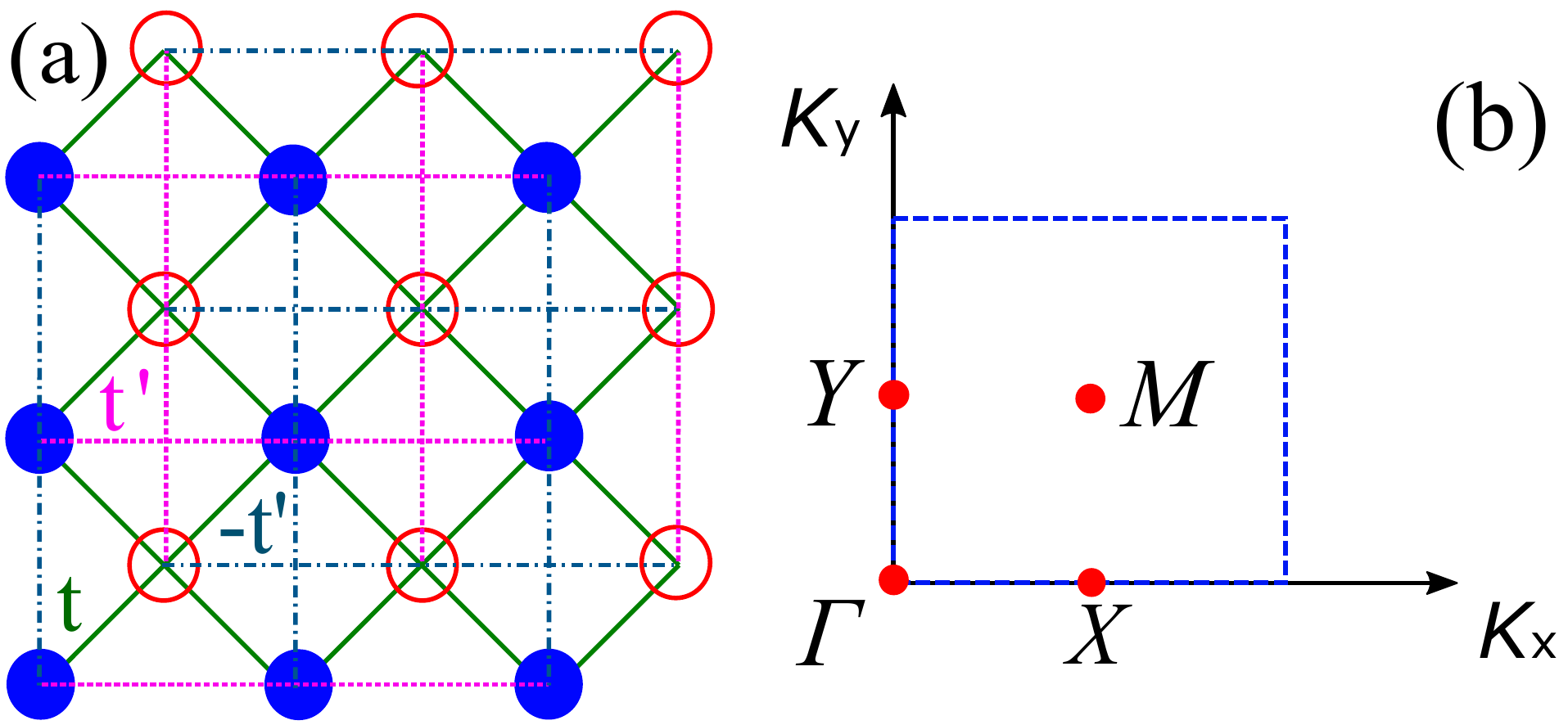}
\caption{Model Hamiltonian and the Brillouin zone of the spinless fermion model on the checkerboard lattice. 
(a) The model on the checkerboard lattice with two sublattices (the solid and empty circles denote the $A$ and $B$ sublattices), which has the nearest-neighbor hopping $t$ and the next-nearest-neighbor hoppings $t'$, $-t'$. 
(b) The Brillouin zone of the model.
Without chemical potential, the non-interacting system at half-filling can realize a semimetal with a quadratic band touching at the $M$ point ${\bs K} = (\pi, \pi)$.
In the presence of on-site energies $\mu / 2$ ($-\mu / 2$) in the $A$ ($B$) sublattice, the quadratic band touching splits to two Dirac band touching points.
The two Dirac band touchings locate either along the line $K_x = \pi$ or the line $K_y = \pi$ depending on the sign of $\mu$.
}
\label{fig:model}
\end{figure}
%%%%%%%%%%%%%%%%%%%%%%%%%%%%%%%%%%%%

In this paper we study a correlated DSM on the checkerboard lattice.
Such a pair of Dirac points may be obtained by applying strain to irradiated FeSe monolayer~\cite{wang2018}, or in certain $\tau$-type organic conductors in the absence of spin-orbit coupling \cite{osada2019} .
A checkerboard lattice may be equivalently considered as a decorated square lattice with two atoms per unit cell, or a bilayer system of two square lattices with the layers  displaced with respect to each other.
Here, we adopt the former perspective, and emphasize the utility of the latter for physical realizations~\cite{wirth2011}. 
For concreteness, we consider a spinless fermion model with a staggered on-site potential for the two sublattices, and repulsive density-density interactions.
The Hamiltonian is given by
\begin{align} 
H &= - \sum_{i, j} ( t_{ij} c_i^{\dagger}c_j + h.c. ) + \frac{\mu}{2} \sum_{i \in A} c_i^{\dagger}c_i - \frac{\mu}{2} \sum_{i \in B} c_i^{\dagger}c_i \nonumber \\
&+ V_1 \sum_{\langle i, j \rangle} n_i n_j + V_2 \sum_{\langle \langle i, j \rangle \rangle} n_i n_j,
\label{eq:ham}
\end{align}
where $t_{ij} = t$ for inter-sublattice hoppings  between nearest-neighbor (NN) `A' and `B' sites, $t_{ij} = t'$ and $-t'$ for intra-sublattice hoppings between NN A (B) sites along the $\hat x$ ($\hat y$) and $\hat y$ ($\hat x$) axes, respectively, as shown in  Fig.~\ref{fig:model}(a).
The strength of the staggered on-site potential is $\mu / 2$, and $V_1$ ($V_2$) is the density-density repulsion between NN A--B (A--A and B--B) sites.
By setting $\mu > 0$ ($\mu < 0$), the two linear or Dirac band-crossing points locate along the line with $K_y = \pi$ ($K_x = \pi$) in the Brillouin zone (BZ). 
In this work we investigate the interaction-induced symmetry broken phases in the particle-hole channel.
We uncover a stable QAH state at small separations between the Dirac points. 
Our results are substantiated by a combination of mean-field calculations and density matrix renormalization group (DMRG) simulation.

This paper is organized as follow.
In Sec.~\ref{sec:dsm}, we describe the DSM phase, and the phase diagram of the Hamiltonian in Eq. \eqref{eq:ham} in the non-interacting limit.
In Sec.~\ref{sec:qah}, we collate all results that support the existence of an interaction-driven QAH state. To this end, we present  mean-field calculations and numerical simulations to show that a QAH state is realized as a finite-coupling instability of the DSM.
We also identify the region in the phase diagram where the QAH phase is expected to be stabilized.
Beyond the regime of stability of the QAH state, a finite interaction strength may lead to other patterns of  symmetry breaking.  
In Sec.~\ref{sec:others}, we discuss non-QAH, symmetry broken states that can directly gap out the Dirac points, and tie the various instabilities to the non-interacting phase diagram. 
A summary and outlook is presented in Sec.~\ref{sec:summary}. 

%%%%%%%%%%%%%%%%%%%%%%%%%%%%%%%%%%%%
%%%%%%%%%%%%%%%%%%%%%%%%%%%%%%%%%%%%

\section{Dirac semimetal phase}
\label{sec:dsm}
In this section we discuss the topology and symmetries of the DSM phases in the non-interacting limit of the model, and deduce the phase diagram as a function of the on-site energy $\mu$.
We define the annihilation and creation operators $(a_{\bf r}, a^{\dagger}_{\bf r})$ and $(b_{\bf r}, b^{\dagger}_{\bf r})$ to denote the fermion operators acting on the  two sites in the unit cell at ${\bf r}$.
It is convenient to formulate the following discussion in the basis of  the two-component fermionic spinor  $\trans{\psi}_{\bs r} = (a_{\bs r} \quad b_{\bs r})$, and introduce a dimensionless parameter for the on-site energy $\delta = \mu / (4t')$.
We note that the non-interacting limit of our model is distinct from the  Mielke model on the checkerboard lattice \cite{mielke1991, montambaux2018, iskin2019}.

In the momentum space the non-interacting, single-particle Hamiltonian takes the form
\begin{equation}
H_0(\bs K) = - d_1(\bs K) \sigma_1 - d_3(\bs K) \sigma_3,
\end{equation}
where $\sigma_j$ is the $j$th Pauli matrix acting on the sub-lattice  degree of freedom, and 
\begin{align}
& d_1(\bs K) = 4t \cos{\frac{K_x}{2}} \cos{\frac{K_y}{2}}, \nn \\
& d_3(\bs K) = 2t' (\cos{K_x} - \cos{K_y} - \delta).
\end{align}
Since $H_0(\bs K)$ is real-valued and composed of even functions of $\bs K$, it is straightforwardly invariant under time reversal operation which acts as $\mc T : \{\bs K \to - \bs K, H_0 \to \mc K H_0 \mc K \}$, where $\mc K$ implements complex conjugation. 
It is also invariant under mirror operations about the $\hat K_x$ and $\hat K_y$ axes passing through $(\pi, \pi)$, which act as $\mathcal M_j: \{K_j \to 2\pi - K_j, H_0 \to \sigma_3 H_0 \sigma_3\}$.
We note that in the limit $\dl \to 0$, the Hamiltonian acquires a fourfold rotational symmetry~\cite{sun2009}.
%%
%%

%%%%%%%%%%%%%%%%%%%%%%%%%%%%%%%%%%%%
\begin{figure}[!t]
\centering
\begin{subfigure}[b]{0.49\columnwidth}
\includegraphics[width=\columnwidth]{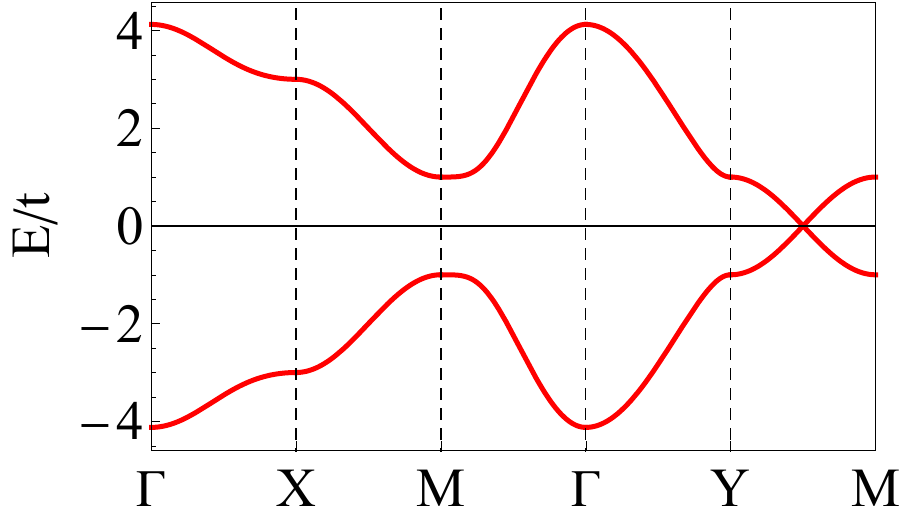}
\caption{$\delta = 1$}
\end{subfigure}
\hfill
\begin{subfigure}[b]{0.49\columnwidth}
\includegraphics[width=\columnwidth]{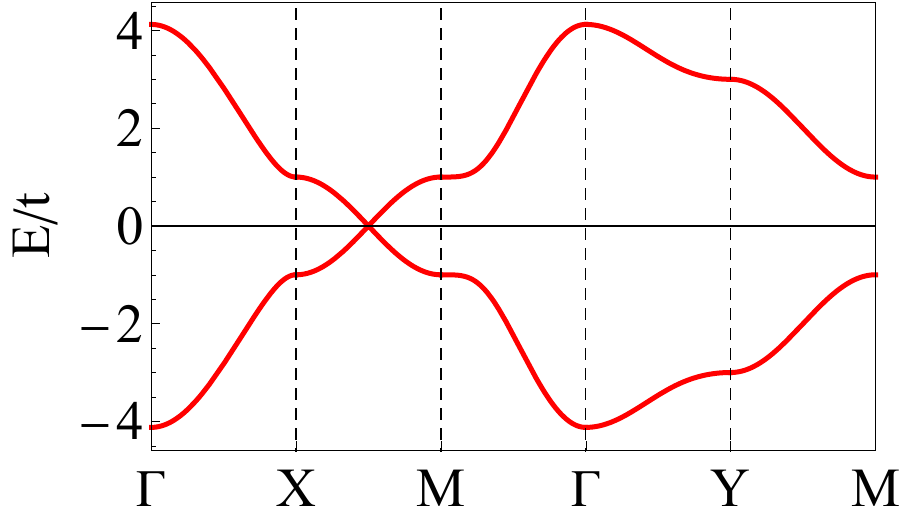}
\caption{$\delta = -1$}
\end{subfigure}
\hfill
\begin{subfigure}[b]{0.49\columnwidth}
\includegraphics[width=\columnwidth]{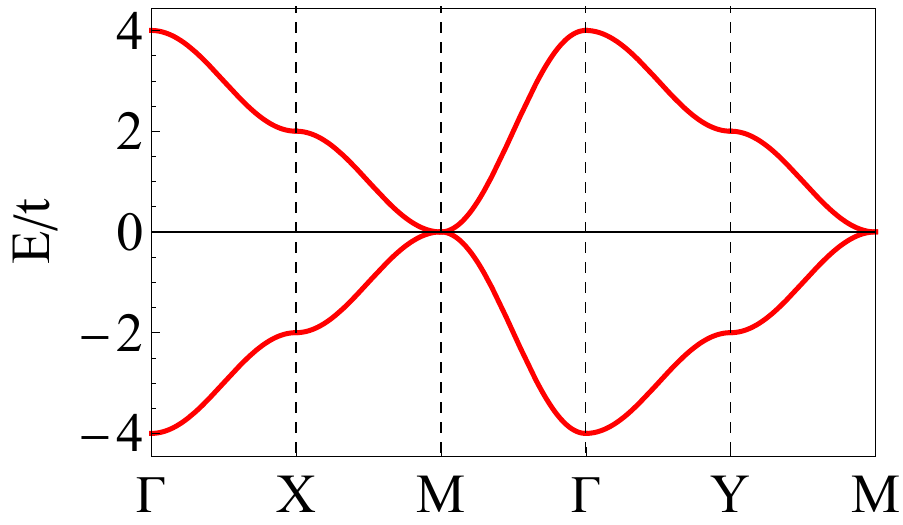}
\caption{$\delta = 0$}
\end{subfigure}
\hfill
\begin{subfigure}[b]{0.49\columnwidth}
\includegraphics[width=\columnwidth]{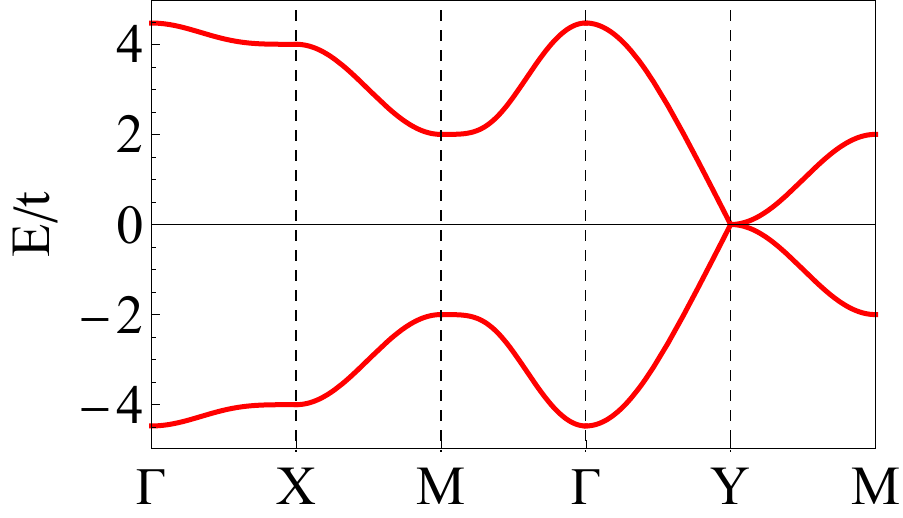}
\caption{$\delta = 2$}
\end{subfigure}
\hfill
\begin{subfigure}[b]{0.6\columnwidth}
\centering
\includegraphics[width=\columnwidth]{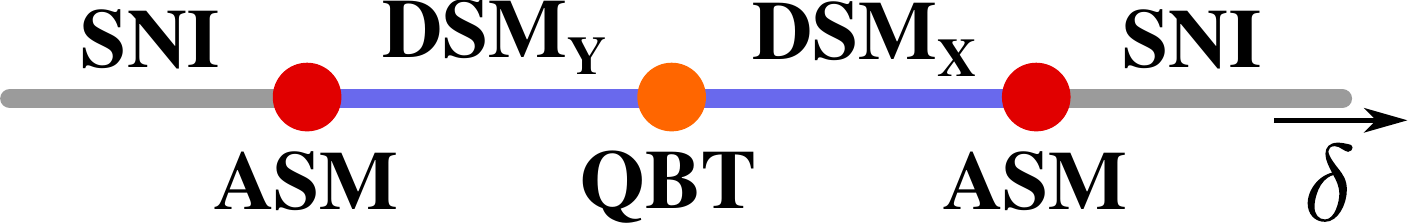}
\caption{}
\label{fig:phase-diag}
\end{subfigure}
\caption{Band structure and phase diagram in the non-interacting limit. (a) $\dl = 1$: a pair of Dirac points are present on the $K_y = \pi$ axis (DSM$_X$ phase). 
(b) $\dl = -1$: a pair of Dirac points are present on the $K_x = \pi$ axis (DSM$_Y$ phase).
At (c) $\dl = 0$ and (d) $\dl =  2$ the points collide to form a quadratic band-touching (QBT) semimetal and an anisotropic semimetal (ASM), respectively.
For $|\dl| > 2$ the system becomes a trivial insulator (SNI), which has different charge densities on the two sublattices.
The $\dl$-driven phase diagram is summarized in (e).
}
\label{fig:bands}
\end{figure}
%%%%%%%%%%%%%%%%%%%%%%%%%%%%%%%%%%%%

For $0 < \delta < 2$ [$-2 < \delta < 0$], a pair of linear band crossings or Dirac points are present at $K_x = \pm \cos^{-1}(-1 + \delta)$ and $K_y = \pi$ [$K_y = \pm \cos^{-1}(-1 - \delta)$ and $K_x = \pi$], as exemplified by Figs. \ref{fig:bands}(a) and \ref{fig:bands}(b).
We call the DSM phase at $\dl > 0$ ($\dl < 0$) DSM$_{\rm X}$ (DSM$_{\rm Y}$).
The linear band crossings in the DSM phases are topologically protected, which is revealed by the winding number along any loop enclosing a single Dirac point.
The winding number along a directed and closed path, $\mc C$, is given by
\begin{align}
W(\mc C) = \frac{1}{2\pi}\oint_{\mc C} \dd{l} ~ \frac{d_1(\bs K) \partial_l d_3(\bs K) - d_3(\bs K) \partial_l d_1(\bs K)}{d_1^2(\bs K) + d_3^2(\bs K)}.
\label{eq:W}
\end{align}
Upon translating $\bs K$ by a reciprocal lattice vector (for example $2\pi \hat K_x$) we find $W(\mc C) \to - W(\mc C)$ due to $d_1(\bs K + 2\pi \hat K_x) = - d_1(\bs K)$.
Since the two Dirac points related by a reciprocal lattice vector cannot carry different vorticities, $W(\mc C) \equiv - W(\mc C)$.
Therefore, $W(\mc C)$ acts as a   $\mathbb Z_2$ index, and can only  distinguish between band singularity points with odd and even vorticities.
Here, we consider $\mc C$ to be directed counter-clockwise, and find that the Dirac points in the DSM phases are characterized by $W(\mc C) = 1$.

At $\delta = 0$ ($\delta = \pm 2$) the Dirac points collide resulting in a QBT semimetal (anisotropic semimetal) where the bands touch at the $M$ point ($Y$ point for $\dl = 2$ and $X$ point for $\dl = -2$), as shown in Figs. \ref{fig:bands}(c) and  \ref{fig:bands}(d).
While $\dl = \pm 2$ are topological quantum critical points that separate DSM phases from ``trivial" insulators \cite{lim2012,isobe2016,sur2019}, the QBT at $\dl = 0$ is a symmetry-protected phase of matter which is protected by a combination of TRS and fourfold rotational symmetry \cite{sun2009}. 
We note that the trivial insulator phase is the site-nematic insulator (SNI) discussed in Ref.~\onlinecite{sur2018}, which has different charge densities on the two sublattices.
Owing to a finite density of states at zero energy, only the QBT at $\dl = 0$ may be destabilized by an arbitrarily weak interaction \cite{sun2009, sur2018}.
In Fig. \ref{fig:bands}(e) we depict  the $\dl$-driven phase diagram for $H_0$.
%%
%%

%%%%%%%%%%%%%%%%%%%%%%%%%%%%%%
\begin{figure}[!t]
\centering
\begin{subfigure}[b]{0.7\columnwidth}
\centering
\includegraphics[width =\columnwidth]{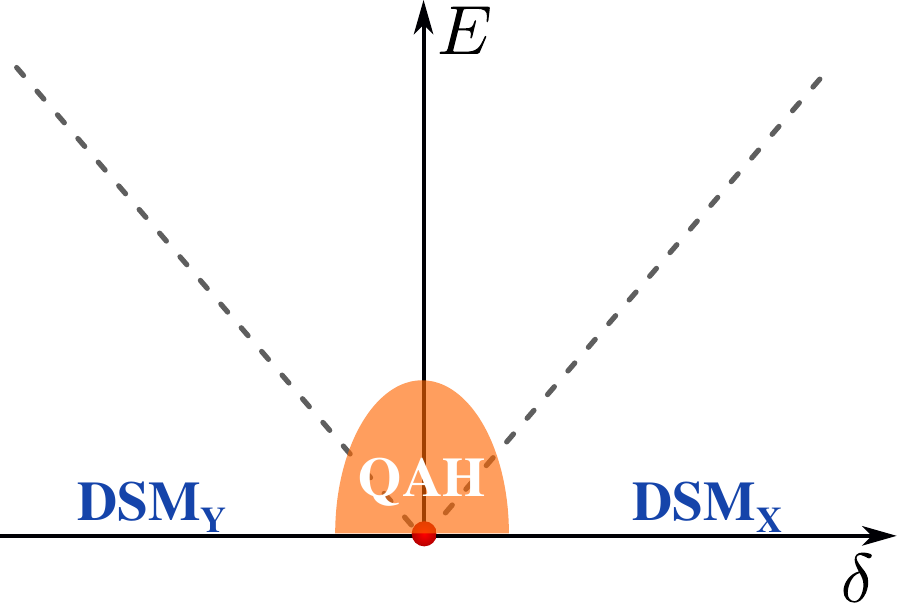}
\caption{}
\label{fig:phase}
\end{subfigure}
\hfill
\begin{subfigure}[b]{\columnwidth}
\centering
\includegraphics[width =\columnwidth]{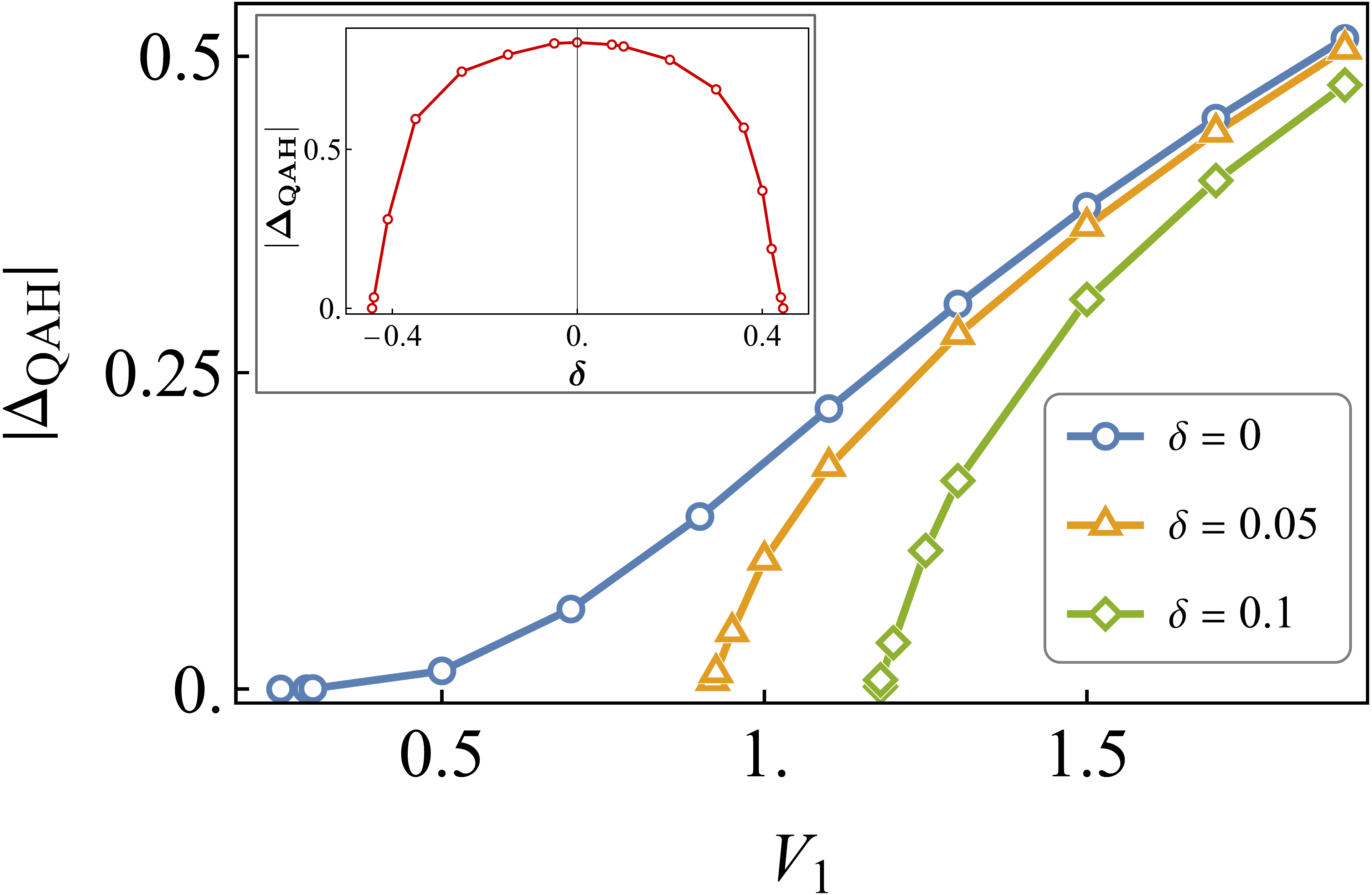}
\caption{}
\label{fig:qah-gap}
\end{subfigure}
\caption{Phase diagram and spectral gap in the quantum anomalous Hall (QAH) state. 
(a) Schematic phase diagram at a fixed interaction strength. The QAH state obtained at $\delta = 0$ extends to a finite region along  $|\delta|$, whose size is controlled by the strength of the interaction.
(b) The behavior of the single-particle excitation gap, $|\Dl_{\rm{QAH}}|$, in the  QAH state as a function of $\dl$ and $V_1$ obtained through a mean-field calculation. While $|\Dl_{\rm{QAH}}| \sim \exp{- 1/V_1}$ at $\dl = 0$, it takes an algebraic form, $|\Dl_{\rm{QAH}}| \sim (V_1 - V_{1c})^\alpha$ with $0<\alpha<1$ and $V_{1c} > 0$, for $|\dl| \neq 0$.
As anticipated in (a), at a fixed $V_1$, $|\Dl_{\rm{QAH}}|$ decreases with increasing $|\dl|$ (see inset), resulting in a dome shaped region about $\dl = 0$ where the QAH phase is stabilized.
In the inset $V_1 = 3.5$.
}
\label{}
\end{figure}
%%%%%%%%%%%%%%%%%%%%%%%%%%%%%%

%%%%%%%%%%%%%%%%%%%%%%%%%%%%%%%%%%%%
%%%%%%%%%%%%%%%%%%%%%%%%%%%%%%%%%%%%
\section{Quantum Anomalous Hall state}
\label{sec:qah}

\subsection{Mean-field analyses}

The density of states vanishes linearly with energy in the DSM phases, which implies that they are stable against short-ranged interactions that are much weaker than $\mu = 4t' \delta$.
Interactions with strength comparable or larger than $\mu$, however, may destabilize the DSM phases, and open spectral gaps at the Dirac points.  
In this section we investigate one of the most interesting symmetry broken states that may result from the DSM phases -- the QAH state.

In the presence of \emph{arbitrarily} weak interactions the QBT semimetal at $\dl = 0$ is unstable against a fourfold symmetric QAH state, that breaks the $\mathcal T$ and both mirror symmetries~\cite{sun2009,sur2018}.
The fourfold rotational symmetry is lost at a finite $\dl$, and the ground state is an DSM for sufficiently weak interactions.
Since $\dl = 0$ is a critical point in the non-interacting limit, it influences the physics at $\dl \neq 0$ through a critical fan that emanates from it~\cite{sachdevBook}, as depicted by the region above the dashed lines in Fig. \ref{fig:phase}.
The boundary between the critical fan and the DSM phases -- the dashed lines --  is set by an energy scale $E_* \sim |\mu|$, which approximately  tracks  the location of the van-Hove points in the single-particle dispersion.
Following the theory of critical phenomena, the physics at energies exceeding $E_*$ is expected to be controlled by the critical point at $\dl =0$~\cite{sachdevBook}.
We take advantage of this influence of the $\dl = 0$ critical point on the phase diagram to look for the QAH state at interaction strengths $|V_n| \gtrsim E_*$.  

On one hand, the reduced symmetry at finite $\delta$ is expected to disfavor the more rotationally symmetric QAH state.
On the other hand, a weak deviation away from $\dl = 0$ is unlikely to offset the free energy gain significantly enough to immediately suppress the QAH order.
Here, we show that the competition between the two tendencies leads to a finite region about $\dl = 0$ where the QAH state survives .
To this end, let us define the QAH order parameter as \cite{sur2018}
\begin{align}
\Dl_{\rm{QAH}}(\bs r) = \gamma_{\bs 0}(\bs r) + \gamma_{\hat x + \hat y}(\bs r) - \gamma_{\hat x}(\bs r) - \gamma_{\hat y}(\bs r),
\end{align}
where $\gamma_{\bs l}(\bs r) \coloneqq i(a_{\bs r}^\dagger b_{\bs r - \bs l} - \mbox{h.c.})$.
In the momentum space it takes the form
\begin{align}
\Dl_{\rm{QAH}}(\bs K) = 4 \sin{\frac{K_x}{2}} \sin{\frac{K_y}{2}} \psi^\dagger(\bs K) \sigma_2 \psi(\bs K).
\end{align}
Thus, at a mean-field level, the single-particle spectrum in the QAH state is gapped.

Since the QAH state results from the condensation of particle-hole pairs on NN $A$ and $B$ sites, a sufficiently strong $V_1$ or $V_2$ can independently drive a QAH instability~\cite{sur2018}.
Although $V_2$ is a formally irrelevant perturbation at the QBT fixed point in a renormalization group sense, it is on a par with the $V_1$ term at the DSM fixed points.
The $V_2$ term, however, contributes to the QAH instability by generating an effective $V_1$ term through quantum fluctuations~\cite{sur2018}.
Therefore, for simplicity, we set $V_2 = 0$, and  perform an explicit mean-field calculation to determine the $\dl$ and $V_1$ dependence of the spectral gap in the QAH state,
\begin{align}
\Delta_{\rm{QAH}} \coloneqq \int \dd{\bs K} \langle  \Dl_{\rm{QAH}}(\bs K) \rangle.
\end{align}
For notational convenience we set $t = 2t' = 1$, and present the details of the calculation in Appendix~\ref{app:mf}.
In Fig.~\ref{fig:qah-gap} we summarize the results.
We find that as $|\dl|$ increases away from $\dl = 0$, progressively stronger coupling is necessary to destabilize the DSM phases, and the critical strength of $V_1$ above which the DSM phases become unstable tracks the boundary of the critical fan.
Furthermore, at a fixed $V_1$, $|\Delta_{\rm{QAH}}|$ decreases with increasing $|\dl|$, which indicates that the QAH instability of the semimetallic states would lead to a dome-shaped region about $\dl = 0$ as shown in Fig.~\ref{fig:phase}.

The mean-field Hamiltonian describing the gapped single-particle excitation in the QAH phase is given by
\begin{align}
H_{\text{QAH}}(\bs K) = H_0(\bs K) +  d_2(\bs K) \sigma_2,  
\end{align}
where $d_2(\bs K) \coloneqq  \Delta_{\text{QAH}} \sin{\frac{K_x}{2}} \sin{\frac{K_y}{2}}$.
The Brillouin zone supports a quantized flux or Chern number in the QAH phase, whose density is given by
\begin{align}
f(\bs K) = \frac{1}{4\pi} \hat d(\bs K) \cdot \frac{\partial \hat d(\bs K)}{\partial K_x} \times \frac{\partial \hat d(\bs K)}{\partial K_y},
\end{align}
with $\hat d = (d_1, d_2, d_3)/\sqrt{\sum_{n=1}^3 d_n^2}$.
Since both $d_1$ and $d_2$ are composed of half-angles, $f(\bs K)$ is periodic under translation of $\bs K$ by a reciprocal lattice vector.
Thus, unlike the vorticity in the non-interacting limit, positive and negative Chern numbers are distinguished in the QAH phase.
Here, the Chern number is found to be $1$.
We note that, owing to the half-angles, the texture of $\hat d$  at any $\dl$ is such that $\hat d$ does not acquire a polar orientation at all high-symmetry locations.
Consequently, the texture wraps only half of the target manifold, $S^2$.
%%
%%

%%%
%%%%%%%%%%%%%%%%%%%%%%%%%%%%%%%%%
\begin{figure}[!t]
\centering
\begin{subfigure}[b]{0.9\columnwidth}
\includegraphics[width = \columnwidth]{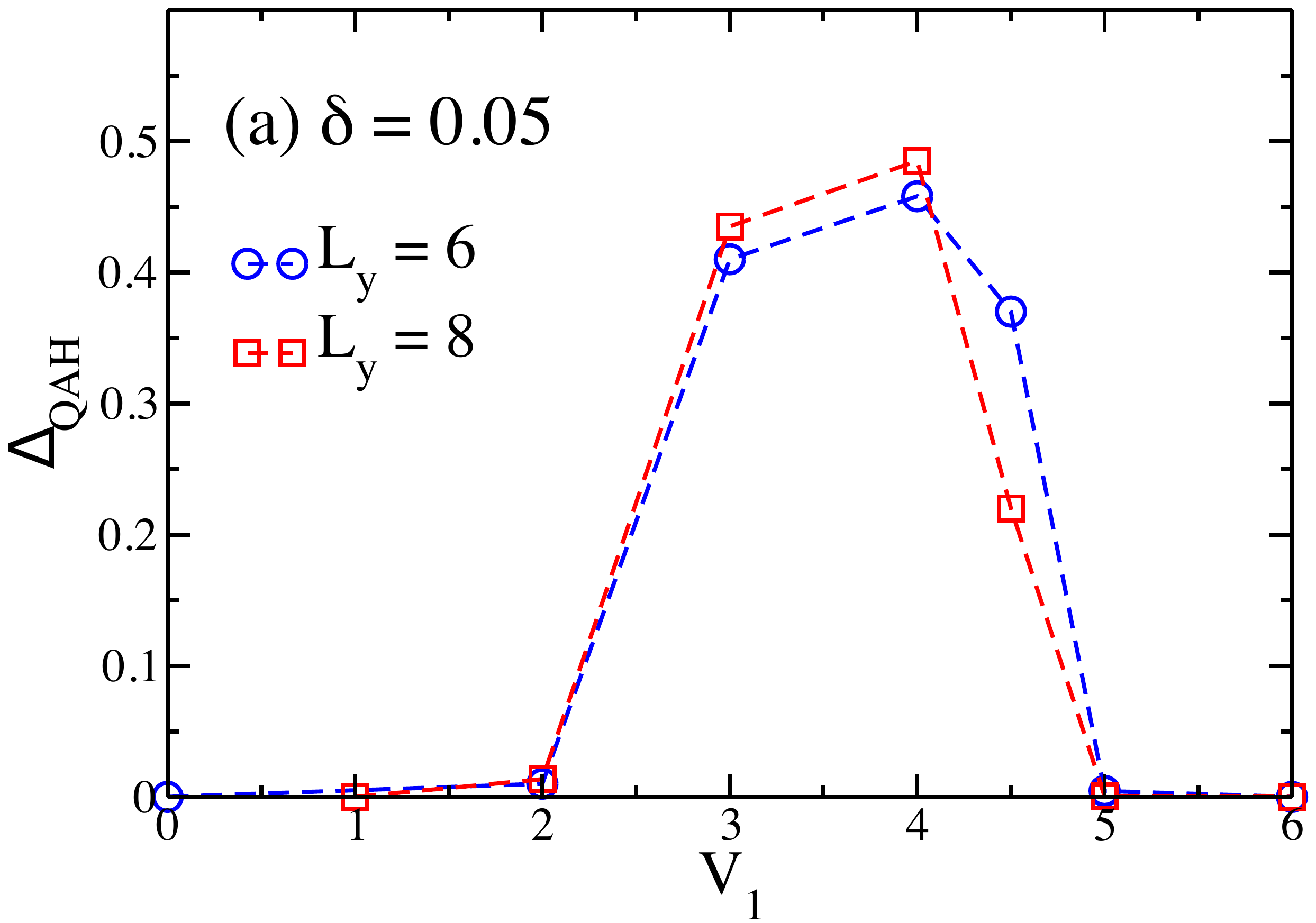}
\caption{}
\end{subfigure}
\hfill
\begin{subfigure}[b]{0.9\columnwidth}
\includegraphics[width = \columnwidth]{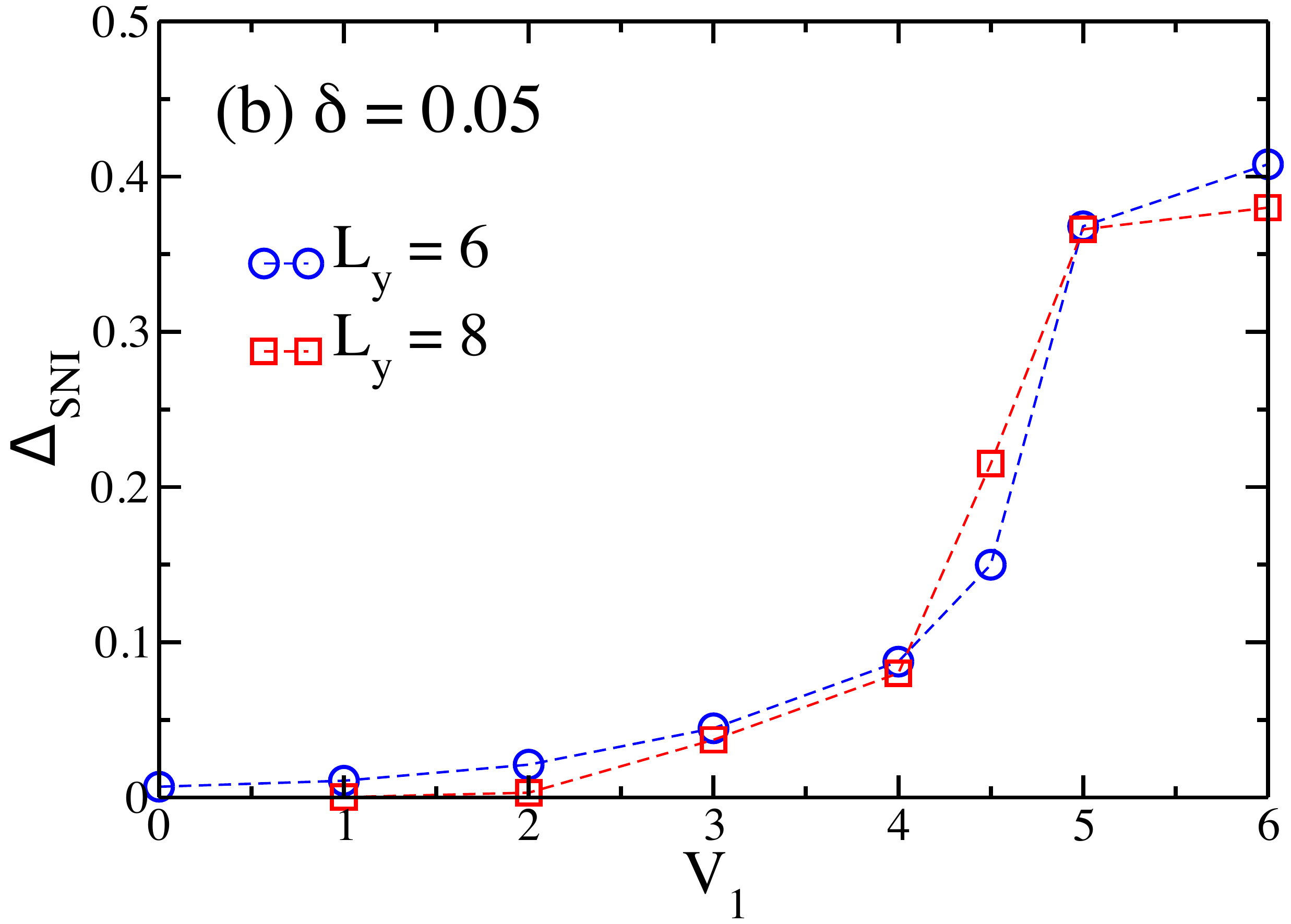}
\caption{}
\end{subfigure}
\hfill
\caption{Identification of the quantum phases induced by interactions.
The non-interacting system is a Dirac semimetal with the on-site energy $\delta = 0.05$.
The repulsive interactions are increased by fixing $V_1 = 2 V_2$.
(a) and (b) are the interaction dependence of the QAH order parameter $\Delta_{\rm QAH}$ and the site-nematic order parameter $\Delta_{\rm SNI}$ on the $L_y = 6, 8$, $L_x = 48$ cylinders, which are obtained by using the bond dimensions $M = 4000$.
The order parameters are measured in the bulk of the cylinders.
}
\label{fig:qah}
\end{figure}
%%%%%%%%%%%%%%%%%%%%%%%%%%%%%%%%%

%%%%%%%%%%%%%%%%%%%%%%%%%%%%%%%%%%%%
%%%%%%%%%%%%%%%%%%%%%%%%%%%%%%%%%%%%

\subsection{Numerical identification of the quantum anomalous Hall state}
\label{sec:numerics1}
Having established the possibility of realizing an QAH state over an extended region of the $\dl$--$V_1$ phase diagram, in this subsection we demonstrate the existence of the QAH state as a finite-coupling instability of the DSM$_{\rm X}$ phase on the checkerboard lattice  by using DMRG simulation~\cite{white1992}.
In particular, we study a cylinder geometry for the system with periodic boundary conditions along the circumference direction ($y$ direction) and open boundary conditions along the axis direction ($x$ direction). 
We use $L_y$ and $L_x$ to denote the numbers of unit cells along the two directions, respectively. 
Our system size is up to $L_y = 8$, and $L_x$ is increased up to $48$ in most calculations. The results are well converged with system length in our calculations. We have also checked the results with increased number of DMRG sweep, confirming all the results converged with sweeping.
We implement the particle number conservation and keep the optimal states up to $M = 4000$ to ensure the truncation error about $1\times 10^{-5}$.
We start from the DSM$_{\rm X}$ phase by choosing a nonzero on-site energy $\delta > 0$ in the Hamiltonian Eq.~\eqref{eq:ham}.
In this case, the Dirac points locate on the $K_y = \pi$ axis but $K_x$ can be incommensurate. While the even $L_y$ is compatible with $K_y = \pi$, the finite-size effects induced by incommensurate $K_x$ can be reduced by increasing system length. 
In Appendix~\ref{app:convergence}, we show the good convergence of the obtained QAH order parameter in the bulk of system versus both system length and bond dimension.

We calculate the QAH order parameter $\Delta_{\rm QAH}$ and the site-nematic order parameter $\Delta_{\rm SNI}$ with growing repulsive interactions.
The site-nematic insulating state has been found in the QBT semimetal in the presence of strong repulsive interactions~\cite{sur2018}.
We define the QAH order parameter for each NN bond $(i, j)$ as $\Delta_{\rm QAH} = 4 i \langle \Psi | c_i^{\dagger}c_j - c_j^{\dagger}c_i | \Psi \rangle$, where $| \Psi \rangle$ is the ground-state wavefunction; and the site-nematic order parameter as $\Delta_{\rm SNI} = |(n_{a} - n_{b})| / 2$, where $n_{a}$ and $n_{b}$ are defined as the particle densities of the two sublattices in the bulk of the system.
To detect possible QAH phase, we consider both the NN $V_1$ and the NNN $V_2$ interactions, which have been found to enhance the QAH order in the QBT semimetal~\cite{zhu2016, sur2018} and may also work in this studied Dirac semimetal.
Otherwise, finite-size DMRG simulation may not be able to identify the QAH order if it is too weak~\cite{sur2018}.
For simplicity, we increase $V_1$ by fixing $V_2 = V_1 / 2$.

In Fig.~\ref{fig:qah}(a), we show the QAH order parameter $\Delta_{\rm QAH}$ with growing interactions obtained by DMRG on the $L_y = 6, 8$ cylinders for $\delta = 0.05$.
We do not show the results for $L_y = 4$, which are vanishing-small due to strong finite-size effects.
To allow spontaneous TRS breaking in DMRG calculation, we choose the wavefunction as complex.
Since DMRG simulation tends to select the minimum entropy state~\cite{jiang2012}, spontaneous TRS breaking is allowed in the complex wavefunction simulation if the energy splitting of the two lowest-energy states are negligible within the resolution of the simulation, because the symmetry breaking states have the minimum entropy. By contrast, DMRG calculation using real wavefunction will obtain a superposition of the two symmetry breaking states, which has a larger entanglement entropy and poses a greater challenge to the convergence of the simulation. Therefore, complex-wavefunction simulations have been widely used for detecting TRS broken states in different systems~\cite{zhu2016, gong2014kagome}.
For $2 \lesssim V_1 \lesssim 4.5$, we find stable nonzero QAH order $\Delta_{\rm QAH}$ in the bulk of the systems, showing the robust spontaneous TRS breaking in this coupling region.
For the NN bonds in the bulk of cylinder, the local QAH ordering pattern results in a loop current that circulates in each plaquette, and the neighboring plaquettes have opposite loop circulation directions, which agrees with a QAH phase with vanished net flux~\cite{haldane1988}.
Here we would like to emphasize that for the smaller interactions, $\Delta_{\rm QAH}$ might be present but is very weak and thus the system size in our calculation cannot detect the order.
As a result, we may take $V_1 \simeq 2$ as the upper bound of the phase boundary between the DSM and QAH phase.
Here, the key result is the identification of the spontaneous TRS breaking with growing interactions.

%%%%%%%%%%%%%%%%%%%%%%%%%%%%%%%%%%%%
\begin{figure}[!t]
\includegraphics[width = 0.9\linewidth]{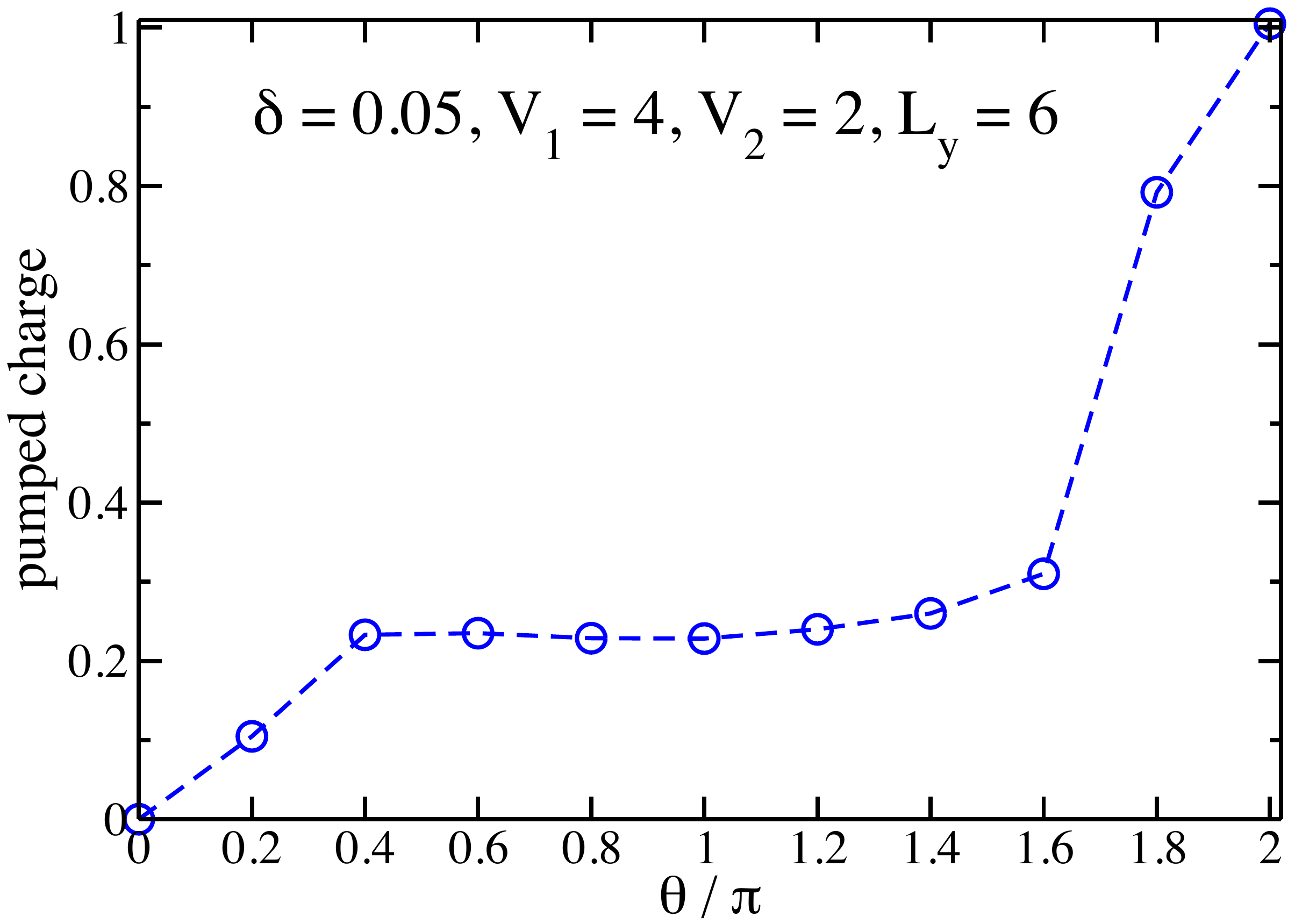}
\caption{Flux insertion simulation in the QAH phase. The flux is adiabatically inserted by using the twisted boundary conditions. The accumulated edge particle number $\delta N$ is obtained by subtracting the particle number in the case of $\theta = 0$. In a period of the flux from $\theta = 0$ to $2\pi$, a quantized charge $\delta N = 1$ is pumped, characterizing a Chern number $C = 1$. In this simulation, a very small additional flux is introduced in all the plaquettes in the purpose of stabilizing a TRS-breaking ground state.
}
\label{fig:chern}
\end{figure}
%%%%%%%%%%%%%%%%%%%%%%%%%%%%%%%%%%%%

In Fig.~\ref{fig:qah}(b), we demonstrate the site-nematic order $\Delta_{\rm SNI}$ versus interactions.
Since the non-interacting system already has a finite on-site potential $\delta$, $\Delta_{\rm SNI}$ must be nonzero in the thermodynamic limit.
In the small-interaction region and the intermediate region with finite $\Delta_{\rm QAH}$, we find that $\Delta_{\rm SNI}$ is small and decreases with increased system circumference $L_y$.
However, $\Delta_{\rm SNI}$ sharply grows for $V_1 \gtrsim 4.5$ accompanied by the vanished $\Delta_{\rm QAH}$, which consistently show a quantum phase transition to a site-nematic insulating phase.

Furthermore, we study the flux response to measure the Hall conductance $\sigma_{\rm H}$ to reveal the topological nature of the intermediate phase~\cite{gong2014kagome, zaletel2014}.
For an IQH state, an integer particle will be pumped from one edge of the cylinder to the other one by adiabatically inserting a period of $U(1)$ flux $\theta$ in the cylinder, following the Laughlin's gedanken experiment~\cite{laughlin1981, sheng2003}.
In a period of flux insertion from $\theta = 0$ to $\theta = 2\pi$, the Hall conductance can be obtained from the pumped particle number $\delta N$ with $\sigma_{\rm H} = \frac{e^2}{h} \delta N$. 
To simulate such a flux insertion in DMRG, we use the twisted boundary conditions in the circumference direction of the cylinder, i.e. $c_i^{\dagger} c_j + h.c. \rightarrow c_i^{\dagger} c_j e^{i \theta} + h.c.$ for all the hopping terms that cross the circumference boundary.
We adiabatically increase the flux $\theta$ in DMRG simulation by using the converged ground state with a given flux $\theta$ as the initial wavefunction for the DMRG sweeping with slightly increased flux $\theta + \delta \theta$~\cite{gong2014kagome, zaletel2014}.
For the converged ground state of each flux $\theta$, we measure the particle density and calculate the accumulated particle number $\delta N$ near the boundaries.
The flux dependence of $\delta N$ is shown in Fig.~\ref{fig:chern}. 
For all the flux values we find no particle accumulation or depletion in the bulk of cylinder.
Only the edge particle accumulation $\delta N$ increases with flux, showing that the particle is pumped from one edge to the other one.
In a period of the flux $\theta = 0 \rightarrow 2\pi$, the pumped particle number is nothing but a precise quantized value $\delta N = 1$, which characterizes this QAH state as a Chern number $C = 1$ IQH state.
Notice that in the previous DMRG study of the interaction-driven QAH state in the semimetals with a quadratic band touching at the Fermi level~\cite{zhu2016, sur2018}, the pumped charge shows a nearly straight line versus $\theta$, indicating the uniform Berry curvature with increased flux~\cite{sheng2006}. However, the charge pumping shown in Fig.~\ref{fig:chern} clearly deviates from a straight line and characterizes the non-uniform Berry curvature with the flux~\cite{sheng2006}.

Here we would like to remark the difficulty in obtaining the quantized Chern number in this DMRG simulation.
Our direct flux insertion simulation does not get $\delta N = 1$ but $\delta N = 0$ at $\theta = 2\pi$, which usually happens when the QAH gap is relatively small compared with the energy splitting between the twofold near-degenerate ground states.
To stabilize the flux insertion simulation, we introduce a very small additional flux in each plaquette, which has been found to be helpful for identifying the QAH state~\cite{zeng2018}.
By using this technique here, we can obtain a quantized Chern number $C = 1$.

%%%%%%%%%%%%%%%%%%%%%%%%%%%%%%%%%%%%
\begin{figure}[t]
\includegraphics[width = 0.9\linewidth]{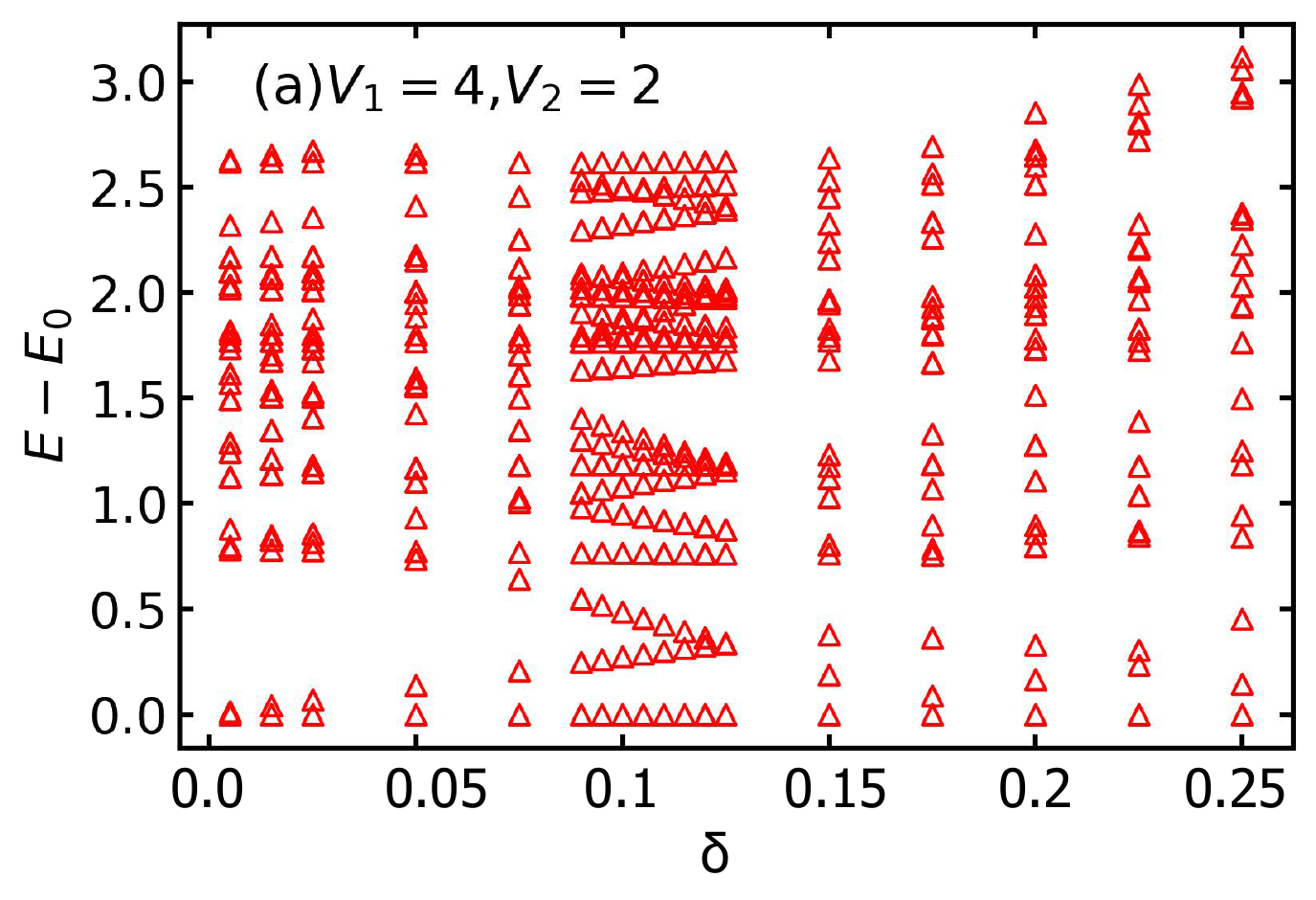}
\includegraphics[width = 0.9\linewidth]{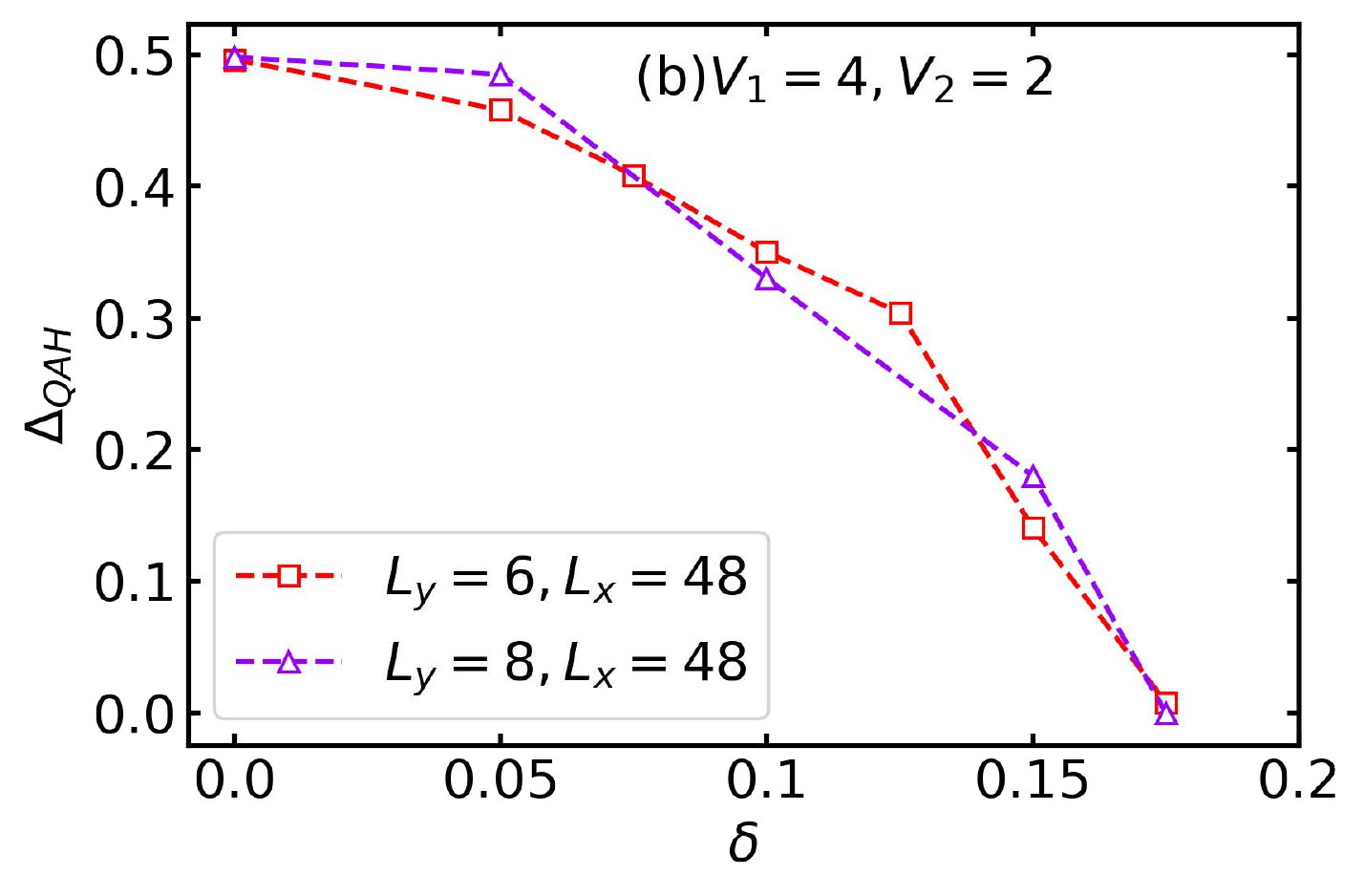}
\caption{Robust QAH phase in the presence of sublattice on-site potential for the model with $V_1 = 4, V_2 = 2$.
(a) Energy spectrum versus the on-site potential $\delta$ on the $L_x = L_y = 4$ torus obtained by ED calculation. For each parameter point $\delta$, all the energy levels have subtracted its ground-state energy $E_0$.
(b) QAH order parameter $\Delta_{\rm QAH}$ versus $\delta$ obtained by DMRG calculation on the $L_y = 6$ and $L_y = 8$ cylinders, which are obtained using the bond dimension $M = 4000$.
}
\label{fig:site}
\end{figure}
%%%%%%%%%%%%%%%%%%%%%%%%%%%%%%%%%%%%

%%%%%%%%%%%%%%%%%%%%%%%%%%%%%%%%%%%%
\subsection{Connection with the quantum anomalous Hall phase in the quadratic band touching semimetal}
\label{sec:numerics2}

We have shown that repulsive interactions can also drive a QAH phase in a Dirac semimetal. 
In this subsection, we unveil that this QAH phase is smoothly connected with the interaction-induced QAH phase in the QBT semimetal, by turning on  anisotropic interactions in the Hamiltonian.  
%Here, we consider two types of anisotropic interactions.
In particular, we focus on the sublattice on-site potential that we have studied.
We start from the QAH phase without on-site potential ($\delta = 0$) at $V_1 = 4, V_2 = 2$, which is induced by interaction in the QBT semimetal~\cite{sur2018}.
By switching on the potential $\delta$, we calculate the energy spectrum of the system on the $L_x = L_y = 4$ torus by using the exact diagonalization (ED).
The potential dependence of energy spectrum is shown in Fig.~\ref{fig:site}(a).
For small potential $\delta$, the nearly double-degenerate ground states which characterize the QAH phase are very robust.  
With growing $\delta$, the gap decreases and closes at $\delta \simeq 0.1$.
In Fig.~\ref{fig:site}(b), we demonstrate the obtained QAH order parameter $\Delta_{\rm QAH}$ by using DMRG.
Based on the DMRG results on the $L_y = 6, 8$ cylinders, we find stable QAH order for $\delta \lesssim 0.15$, showing a robust QAH phase in this region.
The smaller QAH region in the ED results ($\delta \lesssim 0.1$) may be owing to the stronger finite-size effects.

%%%%%%%%%%%%%%%%%%%%%%%%%%%%%%%%%%%%
%%%%%%%%%%%%%%%%%%%%%%%%%%%%%%%%%%%%
\section{Other symmetry broken states}
\label{sec:others}
In Sec.~\ref{sec:qah} we demonstrated the robustness of the interaction induced QAH phase at small $\dl$ and  sufficiently large interaction strengths.
We also argued that this QAH state, in fact, should be considered as an instability of the QBT semimetal, which is inherited by the DSM phases when the interaction strength places the system in the critical fan in the  vicinity of $\dl = 0$. 
In this section, we investigate those symmetry broken states that are true instabilities of the DSM phases, i.e. they gap out single-particle excitations only if $\dl \neq 0$.
For concreteness, we consider the DSM$_{\rm X}$ phase with $\delta > 0$.
The properties of DSM$_{\rm Y}$ can be deduced directly from the results obtained here.

Since the linear dispersion supported by the Dirac points are well-defined only below the van-Hove scale that  $\sim E_*$, we coarse-grain to energies $E \ll E_*$, and focus on the Dirac points  at $\bs K = (\kappa_\pm, \pi)$ where $\kappa_\pm \coloneqq \pi \pm \cos^{-1}(1-\delta)$.
An appropriate description of the low energy dynamics in the vicinity of the Dirac points is formulated in terms of the coarse-grained fermionic operators, $\psi_\pm$, such that 
\begin{equation}
\begin{pmatrix} a_{\bf r} \\ b_{\bf r} \end{pmatrix} \simeq e^{i \kappa_+ x} \psi_{+}({\bf r}) + e^{-i \kappa_- x} \psi_{-}({\bf r}).
\label{eq:psiPM}
\end{equation}
In the basis of the bi-spinor  $\Psi^{\mathsf{T}} = (\psi_+, \psi_-)$, the Hamiltonian, linearized in the vicinity of the Dirac points, obtains the form 
\begin{equation}
h_0(\bs k) =  v_y(\delta) k_y \Gamma_1 + v_x(\delta) k_x \Gamma_{3},
\label{eq:h-lin}
\end{equation}
where $v_x(\delta) = \sqrt{\delta (2 - \delta)}$, $v_y(\delta) = \sqrt{2\dl}$, and ${\bs k} = {\bs K} - {\bs K_D} $ with ${\bs K_D}$ being the location of a Dirac point.
Here, we have defined $(\Gamma_{j}, \Gamma_4, \Gamma_5) = (\tau_3 \otimes \sigma_j, \tau_2 \otimes \sigma_0, \tau_1 \otimes \sigma_0)$ with $j=1, 2, 3$, and $\tau_j$ ($\tau_0$) being the $j$-th Pauli ($2\times 2$ identity) matrix which acts on the valley degree of freedom labeled by `$\pm$' in Eq. \eqref{eq:psiPM}.
Note that we have set $t = 2t' = 1$.
%%

%%%%%%%%%%%%%%%%%%%%%%%%%%%%%%%
\begin{table}[!t]
{\renewcommand{\arraystretch}{1.5}%
\begin{tabular}{| c || c | c |}
\hline 
Symmetry ~&~ Operation ~&~  Broken by   \\ 
\hline \hline
Time reversal & $\bs k \mapsto -\bs k; \; h_0 \mapsto   \Gamma_{4} \mc{K} ~h_0~  \mc{K} \Gamma_{4}$  & $M_4, M_5, M_{13}$ \\
\hline 
$x$-Mirror & $k_x \mapsto - k_x; \; h_0 \mapsto   \Gamma_{34} ~h_0~   \Gamma_{34}$  & $M_{4}, M_{13}$\\
\hline 
$y$-Mirror & $k_y \mapsto - k_y; \; h_0 \to   \Gamma_{12} ~h_0~   \Gamma_{12}$  & $M_{2}, M_{13}$ \\
\hline 
Chiral & $h_0 \to   e^{-i \theta \hat n(\bs k) \cdot  \vec \gamma} ~h_0~  e^{i \theta \hat n(\bs k) \cdot  \vec \gamma}$  & $M_2, M_{4}, M_5$ \\
\hline 
\end{tabular}
}
\caption{The symmetries of the low-energy effective Hamiltonian $h_0$, and the symmetry broken states that open a single-particle excitation gap at the Dirac points. The first column lists the symmetries that protect the DSM$_{\rm X}$ phase. The corresponding symmetry operations on momentum and the Hamiltonian are listed in the second column. Finally, the third column lists the ordering patterns that break the respective symmetries.
The mass terms, $M_{ij}$, are discussed in the main text. 
Here, $\vec \gamma = (\Gamma_{24}, \Gamma_{25}, \Gamma_{45})$, $\theta$ is a real-valued angle, and $\hat n(\bs k)$ is a generically momentum-dependent, 3-component unit vector. 
}
\label{tab:symm}
\end{table}
%%%%%%%%%%%%%%%%%%%%%%%%%%%%%%%

The discrete transformations $\mathcal T$ and $\mathcal M_x$ exchange the two Dirac points, but $\mathcal M_y$ does not.
The $4\times 4$ representations of the microscopic symmetry operations  discussed in Sec. \ref{sec:dsm} are listed in Table~\ref{tab:symm}.
Here, $\Gamma_{mn} \coloneqq  [\Gamma_a, \Gamma_b]/(2i)$.
In addition to the  microscopic symmetries that $h_0$ inherits from $H_0$, the effective Hamiltonian also possesses an emergent $SU(2)$ chiral  symmetry due to the linearization of the dispersion around the Dirac points.
The group of chiral transformations is generated by $\Gamma_{24}$, $\Gamma_{25}$, and $\Gamma_{45}$, with $\Gamma_{45}$ protecting the Dirac points against hybridization.
Thus, the combination of microscopic discrete and emergent chiral symmetries protect the DSM phase.
%%
%%

%%%%%%%%%%%%%%%%%%%%%%%%%%%%%%
\begin{figure}[!t]
\centering
\begin{subfigure}[b]{0.8\columnwidth}
\includegraphics[width=\columnwidth]{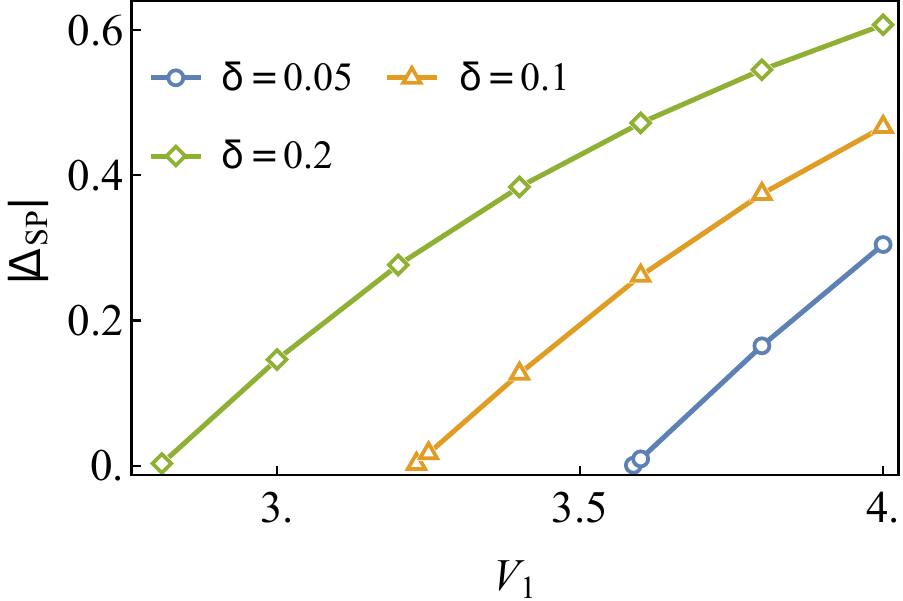}
\caption{}
\label{fig:sp}
\end{subfigure}
\hfill
\begin{subfigure}[b]{0.8\columnwidth}
	\includegraphics[width=\columnwidth]{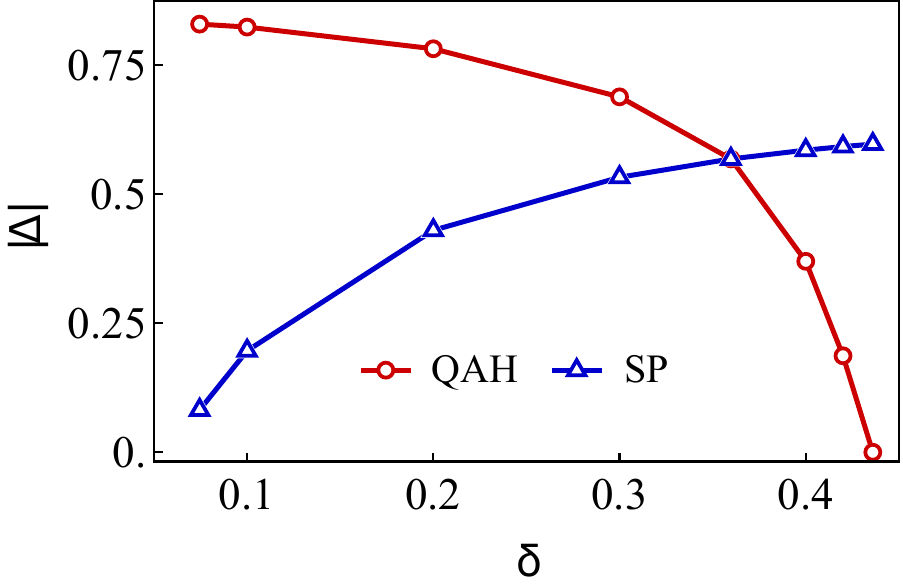}
	\caption{}
	\label{fig:qah-sp}
\end{subfigure}
%\hfill
%\begin{subfigure}[b]{0.8\columnwidth}
%	\includegraphics[width=\columnwidth]{chern-num}
%	\caption{}
%\end{subfigure}
\caption{Competitors of the quantum anomalous Hall (QAH) state. 
(a) The behavior of the mean-field gap in the stripe-Pierls (SP) state. 
Unlike the QAH state, the SP state is strengthened by increasing separation between the Dirac points. 
(b) This leads to an opposite behavior of the gap in the two phases as a function of $\dl$. 
Here, we have fixed $V_1 = 3.5$. 
}
\label{fig:qah-mf}
\end{figure}
%%%%%%%%%%%%%%%%%%%%%%%%%%%%%%

While the vanishing density of states at the Fermi level ensures the stability the DSM phase against weak-coupling instabilities, for sufficiently strong interactions the symmetries that protect the Dirac points may be  spontaneously broken. 
If the resultant symmetry broken state is accompanied by an energy gap in the single particle spectrum, then it would be expected to be stabilized at the cost of the DSM phase.  
Such gap-openings in the particle-hole channel are specified by the $\Gamma$-matrices that anti-commute with $h_0$, viz. $\Gamma_2$, $\Gamma_4$, $\Gamma_5$, and $\Gamma_{13}$.
These correspond to the order parameters $M_j({\bf k}) \equiv \Psi^\dagger({\bf k}) \Gamma_j \Psi({\bf k})$ with $j = 2,4,5$ and $M_{13}({\bf k}) \equiv \Psi^\dagger({\bf k}) \Gamma_{13} \Psi({\bf k})$.
The symmetries broken by individual mass orders are listed in Table~\ref{tab:symm}.
While $M_2$ and $M_{13}$ involve intra-valley particle-hole order, $M_4$ and $M_{5}$ hybridize the two Dirac points.
In particular, $M_{13}$ corresponds to the QAH order parameter, and the QAH mass gap takes the same sign at the two Dirac points, indicating its source to be different than the DSM phase. 
By contrast, the non-QAH orders are sensitive to the existence of the Dirac points, either through a sign-change of the mass gap ($M_2$),  or spatial modulation over a scale $\sim (\kappa_+ - \kappa_-)^{-1}$ ($M_4$ and $M_5$).

In Sec.~\ref{sec:qah} we have shown that the QAH order is progressively suppressed by increasing separation between the Dirac points.
How do the other symmetry-broken states fare compared to the QAH state?
By focusing only on patterns of symmetry breaking that are commensurate with the checkerboard lattice, we compare the behaviors of the QAH state and the state with $M_2$ as the order parameter.
The latter is characterized by Pierls-like distortion along the $\hat x$ direction that modulates over a single unit-cell spacing.
Consequently, we call it  ``stripe-Pierls" (SP) state.
In contrast to the QAH state (see Fig.~\ref{fig:qah-gap}), the gap in the SP state increases with both $\dl$ and $V_1$ as shown in Fig.~\ref{fig:sp}.
This opposite tendency of the gaps in the two symmetry-broken states, as a function of $\dl$,  suggests that at a fixed interaction strength, for a sufficiently large separation between the Dirac points, the SP state would  eventually dominate over the QAH state.
We demonstrate it by plotting the behavior of the respective gaps as a function of $\dl$ in Fig.~\ref{fig:qah-sp}.
Thus, it would be expected that the QAH phase would give way to the SP phase at sufficiently large separation between the Dirac points.
%%
%%

%%%%%%%%%%%%%%%%%%%%%%%%%%%%%%%%%%%%
\begin{figure}[!t]
\includegraphics[width = 0.9\linewidth]{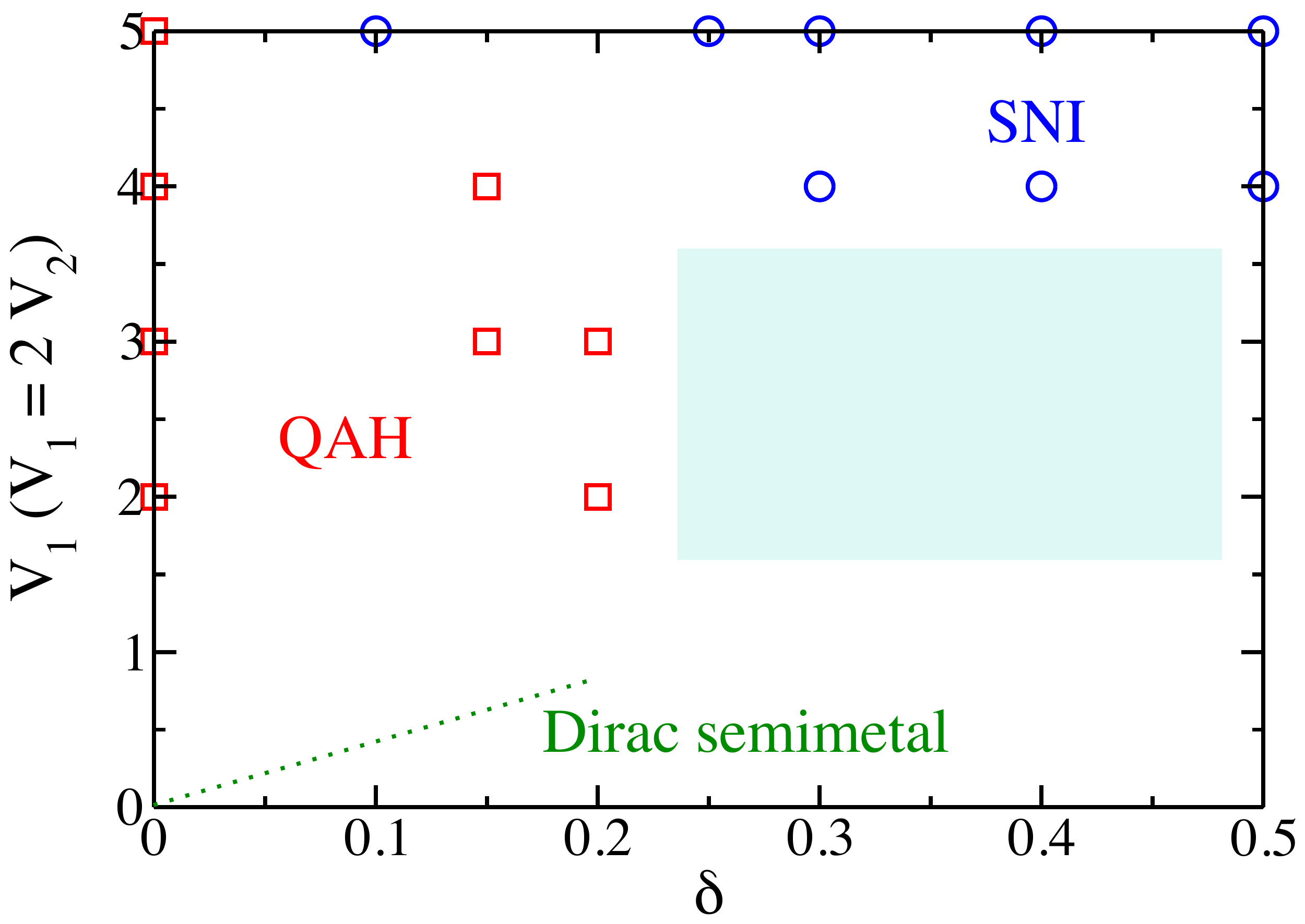}
\caption{Schematic phase diagram of the model with increased on-site energy $\delta$ and interactions.
Here we set $V_1 = 2V_2$. The symbols denote the parameter points which we determine their orders by DMRG calculation, with the red squares for the QAH state and the blue circles for the SNI state. For small interactions, the Dirac semimetal would be stable. The dotted line is a schematic phase boundary between the QAH and semimetal. With growing interactions in the larger-$\delta$ regime (the colored regime), DMRG results do not find evidence to support a symmetry breaking phase before the SNI state appears.
}
\label{fig:phaseDMRG}
\end{figure}
%%%%%%%%%%%%%%%%%%%%%%%%%%%%%%%%%%%%

%%
We numerically explore the quantum phase diagram of the system with growing $\delta$ and interactions by choosing $V_1 = 2V_2$.
In the DMRG calculation on the $L_y = 6$ systems, we identify the QAH and the SNI state by computing the corresponding order parameters, as we have done in Fig.~\ref{fig:qah}.
The obtained results are shown as the symbols in Fig.~\ref{fig:phaseDMRG}.
For small interactions, the Dirac semimetal would be stable against interactions.
The dotted line depicts a schematic phase boundary between the QAH and Dirac semimetal.
With increased interactions in the larger-$\delta$ regime (the colored regime in Fig.~\ref{fig:phaseDMRG}), our DMRG calculations do not find evidence to support either bond order or site order before the SNI state emerges (see the results in Appendix~\ref{app:suppl_data}), indicating a direct transition from the DSM to the SNI phase.
This is a resonable possibility because 
% no symmetry protects the  absolute location of the Dirac points along the $\hat K_x$ axis, and $\delta$ is subject to renormalizations. 
% %%
% In particular, 
the DSM phase may be considered as a state arises out of a nematic instability of the QBT semimetal.
Previous calculations suggest that the corresponding nematic order, $\langle \psi^\dagger(\bs K) \sigma_3 \psi(\bs K) \rangle$, is strengthened by increasing repulsive interactions~\cite{sur2018}, which would translate into  $|\dl|$ renormalizing to larger values.
Larger $|\delta|$, however,  pushes the DSM towards an SNI state, as illustrated in Fig.~\ref{fig:phase-diag}.
Therefore, it is possible that repulsive interactions induce a direct transition from the  DSM phase to the SNI phase at a  sufficiently strong  bare $|\delta|$. 
% without a discernible presence of the SP phase. 
%%
We note that, although our DMRG simulations suggest an absence of the SP phase as the QAH state is suppressed, it may not rule out incommensurate ordering patterns, eg. $M_4$ and $M_5$, due to the limit of system size in numerical simulation. 
%% 
%%

%%%%%%%%%%%%%%%%%%%%%%%%%%%%%%%%%%%%
%%%%%%%%%%%%%%%%%%%%%%%%%%%%%%%%%%%%
\section{Conclusion}
\label{sec:summary}
In this work we considered spinless fermions hopping on a checkerboard lattice under uniaxial strain or staggered on-site potential to   demonstrate that (i) it is indeed possible to realize QAH states in Dirac semimetals by spontaneous TRS breaking; (ii) single-particle quantum  critical points play a fundamental role in guiding the strong-coupling symmetry broken phases in its vicinity.
Through mean-field calculations we argued that the QAH state should be considered as an instability of the quadratic band-touching semimetal  at $\dl = 0$, which survives in the DSM phase at sufficiently strong interactions.
Our numerical simulations support this perspective, and we showed that the stable QAH state at finite $\delta$ is smoothly connected to that at $\delta = 0$.
The suppression of the QAH order with increasing bare $\delta$ provides further support.
We have also identified other symmetry breaking channels that can open a spectral gap at the Dirac points, and potentially competes with QAH state at sufficiently large bare separation between the Dirac points. 
A comprehensive understanding of their mutual competition would require a detailed renormalization group analysis and numerical simulations, both of which are beyond the scope of this work, but would be an interesting topic for future investigations.

Since the spinful version of our model at $\delta = 0$ is  unstable to a quantum spin Hall (QSH) state~\cite{sun2008}, we expect that the corresponding spinful DSM phase obtained by applying uniaxial strain or staggered onsite potential would continue to be unstable to the QSH state at small $\delta$.
In both spinless and spinful models it would also be interesting to investigate potential  pairing instabilities and their relationship with anomalous/spin Hall fluctuations.
We leave such consideration to future work. 
%%

%%%%%%%%%%%%%%%%%%%%%%%%%%%%%%%%%%%%
%%%%%%%%%%%%%%%%%%%%%%%%%%%%%%%%%%%%

\begin{acknowledgments}
This work was supported by the RGC of Hong Kong SAR of China (Grants No. 17303019, No. 17301420 and AoE/P-701/20) (H.Y.L.), the National Natural Science Foundation of China Grants No. 11874078 and No. 11834014 (S.S.G.). 
S.S. was supported by the start up funds of Pallab Goswami provided by Northwestern University, the National Science Foundation MRSEC program (DMR-1720319) at the Materials Research Center of Northwestern University, and the U.S. Department of Energy, Computational Materials Sciences (CMS) program under Award Number DE-SC0020177 at Rice University. 
D.N.S. was supported by National Science Foundation through the Partnership in Research and Education in Materials Grant  DMR-1828019.
\end{acknowledgments}

%%%%%%%%%%%%%%%%%%%%%%%%%%%%%%%%%%%%

\paragraph*{} H.Y.L. and S.S. contributed equally to this work.

\clearpage
\newpage 

\appendix
\onecolumngrid

\section{Mean-field calculations} \label{app:mf}
Here we present the details of the mean-field calculations.
We use the `power expanded Gibbs potential method' (PEGP) \cite{sur2018} for comparing the two commensurate symmetry broken states, viz. quantum anomalous Hall (QAH) and stripe-Pierls (SP), for $\delta>0$.  
For the symmetry breaking channels the action in the presence of the respective source terms are
\begin{align}
& S_{QAH} = S_0^{(QAH)} + S_{int} \label{eq:a3}\  \\
&S_{SP} = S_0^{(SP)} + S_{int},\label{eq:a4}\ 
\end{align}
where 
\begin{align}
S_0^{(QAH)} &= S_0+\int \frac{dk_0}{2\pi}J_{QAH}\Delta_{QAH}\quad 
\\ &=\int dk \ \psi_k^\dagger\ [ ik_0\sigma_0 + 4t \cos{\frac{k_x}{2}}\cos{\frac{k_y}{2}}\sigma_1 + \{2t'(\cos{k_y}-\cos{k_x}) + \mu \}\sigma_3 + 4J_{QAH}\sin{\frac{k_x}{2}}\sin{\frac{k_y}{2}}\sigma_2 ]\,\psi_k \ .
\label{eq:a5} \\
S_0^{(SP)} &= S_0 +\int \frac{dk_0}{2\pi} J_{SP}\Delta_{SP}=\int dk\, \psi_k^\dagger[ik_0\sigma_0 + d_1(\vec k)\sigma_1+d_3(\vec k)\sigma_3 + d_{SP}(\vec k)J_{SP}\sigma_2]\,\psi_k \ . \label{eq:a7}
\end{align}
with $k_0$ being the Euclidean frequency,  $d_1=4t\cos{\frac{k_x}{2}}\cos{\frac{k_y}{2}}$, $d_3=2t'(\cos{k_y}-\cos{k_x})+\mu$, $d_{QAH}(\vec k) = 4\sin{\frac{k_x}{2}}\sin{\frac{k_y}{2}}$, and $d_{SP}(\vec k) = 4\cos{\frac{k_x}{2}}\sin{\frac{k_y}{2}}$.
Note that we have assumed $J_X$ with $X = QAH, SP$ to be independent of $k_0$ and $\vec k$.
Thus, the respective propagators are
\begin{align} 
& G_0^{(QAH)}(k,J_{QAH}) = \frac{-ik_0\sigma_0 + d_1(\vec k)\sigma_1 + d_3(\vec k)\sigma_3 + d_{QAH}(\vec k)\sigma_2J_{QAH}}{ k_0^2 + d_1^2(\vec k) + d_3^2(\vec k) +J_{QAH}^2d_{QAH}^2(\vec k) }, \label{eq:a6} \\
& G_0^{(SP)}(k,J_{SP})=\frac{-ik_0\sigma_0 + d_1\sigma_1 + d_3\sigma_3 + J_{SP}d_{SP}\sigma_2}{k_0^2 + d_1^2(\vec k) + d_3^2(\vec k) + d_{SP}^2(\vec k)J_{SP}^2}. \label{eq:a8}
\end{align}
In either case
\begin{equation}
\begin{aligned}\label{eq:a9}
S_{int} =& \ 4V_1 \int dk_1 dk_2 dq\ \cos{\frac{q_x}{2}} \cos{\frac{q_y}{2}}\ a^{\dagger}(k_1+q)a(k_1)b^{\dagger}(k_2)b(k_2+q)\\
& +2V_2\int dk_1dk_2dq\ [(\sin{\frac{k_{1x}-k_{2x}}{2}}\sin{\frac{k_{1x}-k_{2x}+q_x}{2}} )+ (x\rightarrow y)]\\ & \times[(a^{\dagger}(k_1+q)a(k_1)a^{\dagger}(k_2-q)a(k_2) )+ (a \rightarrow b)]
\end{aligned}
\end{equation}
We note that the respective order parameters and sources are related by 
\begin{align}
& \Delta_{QAH}= \langle \int d\vec k\  4\sin{\frac{k_x}{2}}\sin{\frac{k_y}{2}}\psi_k^\dagger\sigma_2\psi_k \rangle=-4\int d\vec k \sin{\frac{k_x}{2}}\sin{\frac{k_y}{2}}\ {\rm Tr}[\sigma_2G_0^{(QAH)}(k)]\ , \label{eq:a1} \\
& \Delta_{SP}= \langle \int d\vec k\ 4\cos{\frac{k_x}{2}}\sin{\frac{k_y}{2}}\psi_k^\dagger\sigma_2\psi_k \rangle=-4\int d\vec k \cos{\frac{k_x}{2}}\sin{\frac{k_y}{2}}\ {\rm Tr}[\sigma_2G_0^{(SP)}(k)]. \label{eq:a2}
\end{align}

According to the PEGP expansion, up to linear order in interaction strength, the Gibbs free energy $\mathcal{G}(\Delta)=\mathcal{G}_0(\Delta)+\langle S_{int} \rangle$. 
For the reason noted in the main text henceforth we set $V_2 = 0$, and evaluate $\langle S_{int} \rangle$ individually for either the QAH or SP state, 
\begin{equation}\label{eq:a10}
\begin{aligned}
\langle S_{int} \rangle_X = 4V_1\int dk_1 dk_2 dq\ \cos{\frac{q_x}{2}} \cos{\frac{q_y}{2}}\ [&\langle a(k_1)a^{\dagger}(k_1+q) \rangle_X \langle b(k_2+q)b^{\dagger}(k_2) \rangle_X \\ &- \langle b(k_2+q)a^{\dagger}(k_1+q) \rangle_X \langle a(k_1)b^{\dagger}(k_2) \rangle_X],
\end{aligned}
\end{equation}
where $\ \langle a(k)b^{\dagger}(k')\rangle_X=(2\pi)^3\delta^{(3)}(k-k')G_{ab}^{(X)}(k)$; $\ \langle a(k)a^{\dagger}(k')\rangle _X =(2\pi)^3\delta^{(3)}(k-k')G_{aa}^{(X)}(k)$, etc. with $X$ referring to QAH or SP.
Upon further evaluation we obtain
\begin{equation}
\begin{aligned}\label{eq:a13}
\langle S_{int} \rangle_X = &-4V_1(2\pi)^3\delta^{(3)}(0)[\{ \frac{1}{2}\int d\vec k\frac{d_3(\vec k)}{| M(\vec k) |} \}^2
\\ &+\frac{1}{4}\int d\vec k_1 d\vec k_2 \frac{d_1(k_1)d_1(k_2)+J_X^2d_X(k_1)d_X(k_2)+iJ_X\{ d_1(k_1)d_X(k_2)-d_1(k_2)d_X(k_1) \}}{|M(k_1)|\ |M(k_2)|}\\ &\quad\times \{ \cos{\frac{k_{1x}}{2}}\cos{\frac{k_{2x}}{2}}\cos{\frac{k_{1y}}{2}}\cos{\frac{k_{2y}}{2}}+\cos{\frac{k_{1x}}{2}}\cos{\frac{k_{2x}}{2}}\sin{\frac{k_{1y}}{2}}\sin{\frac{k_{2y}}{2}}
\\&\qquad+\sin{\frac{k_{1x}}{2}}\sin{\frac{k_{2x}}{2}}\cos{\frac{k_{1y}}{2}}\cos{\frac{k_{2y}}{2}}+\sin{\frac{k_{1x}}{2}}\sin{\frac{k_{2x}}{2}}\sin{\frac{k_{1y}}{2}}\sin{\frac{k_{2y}}{2}} \}]
\end{aligned}
\end{equation}

\begin{itemize}
\item For QAH state: At $\mu=0$, $M(\vec k)$ is invariant under $k_x\longleftrightarrow k_y$, but $d_3(\vec k)$ is odd. 
Thus, the 1st term in Eq.~\eqref{eq:a13} vanishes.
At $\mu \ne 0$, the above is no longer true and the Hartree term contributes.
At any $\mu$, $M(\vec k)$ is invariant under $k_j \rightarrow 2\pi-k_j$, but $d_1(\vec k)$ is odd.
Thus, the non-vanishing terms in the Fock term are (numerator only): $$d_1(\vec k_1)d_1(\vec k_2)\cos{\frac{k_{1x}}{2}}\cos{\frac{k_{1y}}{2}}\cos{\frac{k_{2x}}{2}}\cos{\frac{k_{2y}}{2}} + J_{QAH}^2d_{QAH}(\vec k_1)d_{QAH}(\vec k_2)\sin{\frac{k_{1x}}{2}}\sin{\frac{k_{1y}}{2}}\sin{\frac{k_{2x}}{2}}\sin{\frac{k_{2y}}{2}}.$$
Therefore:
\begin{equation}
\label{eq:a14}
\langle S_{int} \rangle_{QAH}=-V_1 (2\pi)^3\delta^{(3)}(0) [f_3^2+f_1^2+J_{QAH}^3f_{QAH}^2]
\end{equation}
where $f_3(J_{QAH})=\int d\vec k\frac{d_3(\vec k)}{M(\vec k)}$, $f_1(J_{QAH})=\int d\vec k\frac{d_1^2(\vec k)}{4tM(\vec k)}$, and $f_{QAH}(J_{QAH})=\int d\vec k \frac{d_{QAH}^2(\vec k)}{4M(\vec k)}$.
\item For SP state the non-vanishing terms in the numerator of the Fock term is 
$$d_1(\vec k_1)d_1(\vec k_2)\cos{\frac{k_{1x}}{2}}\cos{\frac{k_{1y}}{2}}\cos{\frac{k_{2x}}{2}}\cos{\frac{k_{2y}}{2}} + J_{SP}^2d_{SP}(\vec k_1)d_{SP}(\vec k_2)\cos{\frac{k_{1x}}{2}}\sin{\frac{k_{1y}}{2}}\cos{\frac{k_{2x}}{2}}\sin{\frac{k_{2y}}{2}}.$$
Hence,
\begin{equation}
\label{eq:a15}
\langle S_{int} \rangle_{SP}=-V_1 (2\pi)^3\delta^{(3)}(0) [\{ \int d\vec k\frac{d_3(\vec k)}{M(\vec k)} \}^2 +
\{ \int d\vec k\frac{d_1^2(\vec k)}{4tM(\vec k)} \}^2 + J_{SP}^2\{ \int d\vec k\frac{d_{SP}^2(\vec k)}{4M(\vec k)} \}^2]
\end{equation}
\end{itemize}
The Gibbs free energy is extremized  with respect to $\Delta_X$, and the solution for $\partial_{\Delta_X} \mathcal{G}(\Delta_X) = 0$ yields the gap in the symmetry broken state $X$.
We note that, while extremizing, $\partial_{\Delta_X}J_X$ is obtained as a function of $\Delta_X$ by inverting the respective relationships in Eqs. \eqref{eq:a1} and \eqref{eq:a2}.
For more details, we direct the interested reader to Ref.~\onlinecite{sur2018}.
\section{Convergence of the quantum anomalous Hall order parameter} \label{app:convergence}

\begin{figure}[t]
\includegraphics[width = 0.48\linewidth]{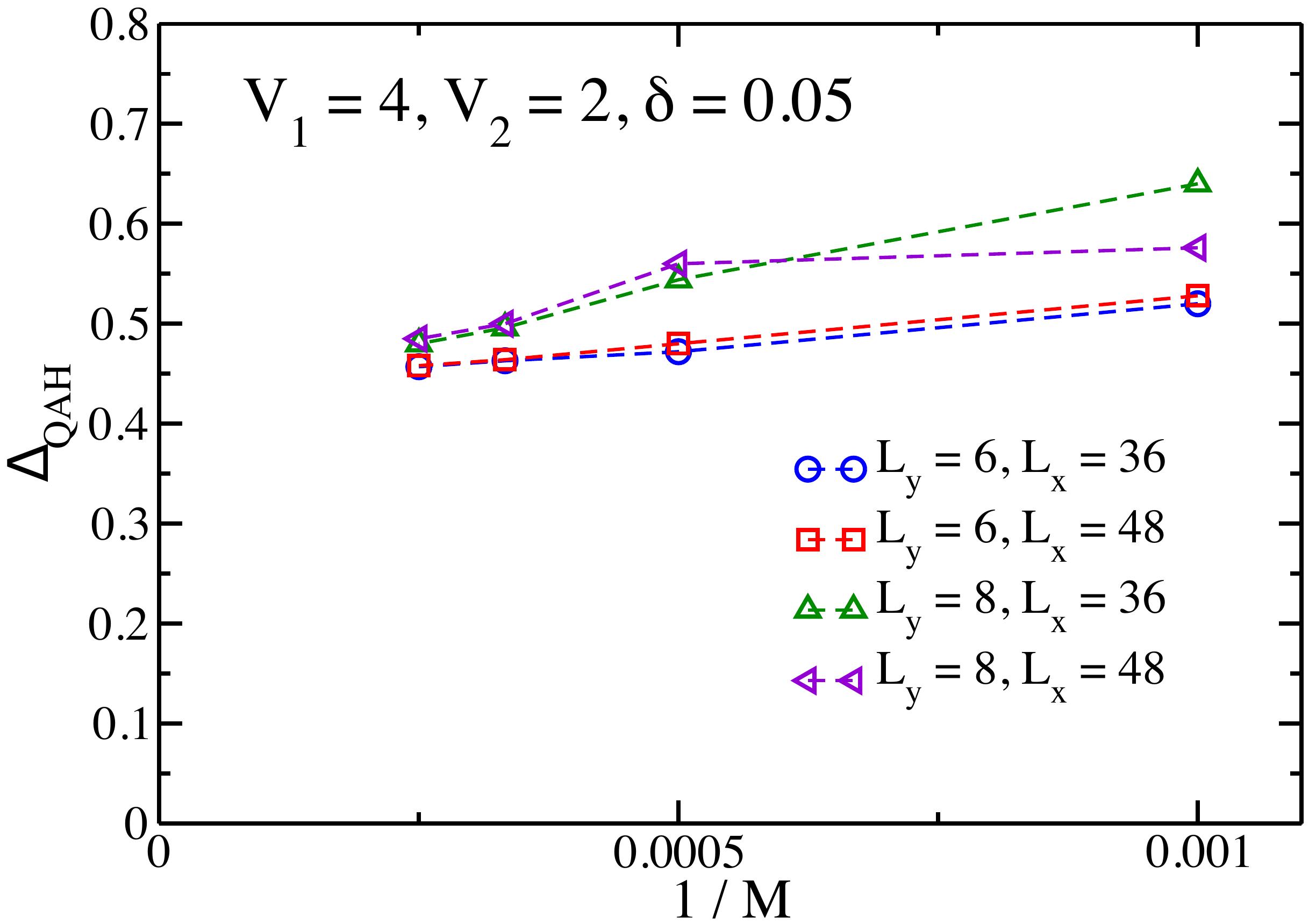}
\caption{Bond dimension dependence and system length dependence of the QAH order parameter. The system has $V_1 = 4, V_2 = 2, \delta = 0.05$ on the $L_y = 6$, $L_x = 36, 48$ cylinders and the $L_y = 8$, $L_x = 36, 48$ cylinders. $M$ is the bond dimension, which is kept from $M = 1000$ to $M = 4000$ in our simulation. 
}
\label{fig:convergence}
\end{figure}

In the main text, we have shown the QAH order parameter on the $L_x = 48$ cylinder.
In our DMRG calculation, we have checked the convergence of the QAH order versus bond dimension $M$ and system length $L_x$.
Here, we present the bulk QAH order on the $L_y = 6, 8$ cylinders with different system lengths $L_x = 36, 48$, which are obtained by using complex wavefunction and keeping $M = 1000 - 4000$ states.
The results for $V_1 = 4, V_2 = 2, \delta = 0.05$ are shown in Fig.~\ref{fig:convergence}.
One can find that the bulk QAH order is well converged with growing system length and the results by keeping $4000$ states are convergent, which supports the accuracy of our DMRG results.

%%%%%%%%%%%%%%%%%%%%%%%%%%%%%%%%%%%%
%%%%%%%%%%%%%%%%%%%%%%%%%%%%%%%%%%%%

%%
\section{More measurements for the larger potential regime} \label{app:suppl_data}

\begin{figure}[t]
\includegraphics[width = 0.9\linewidth]{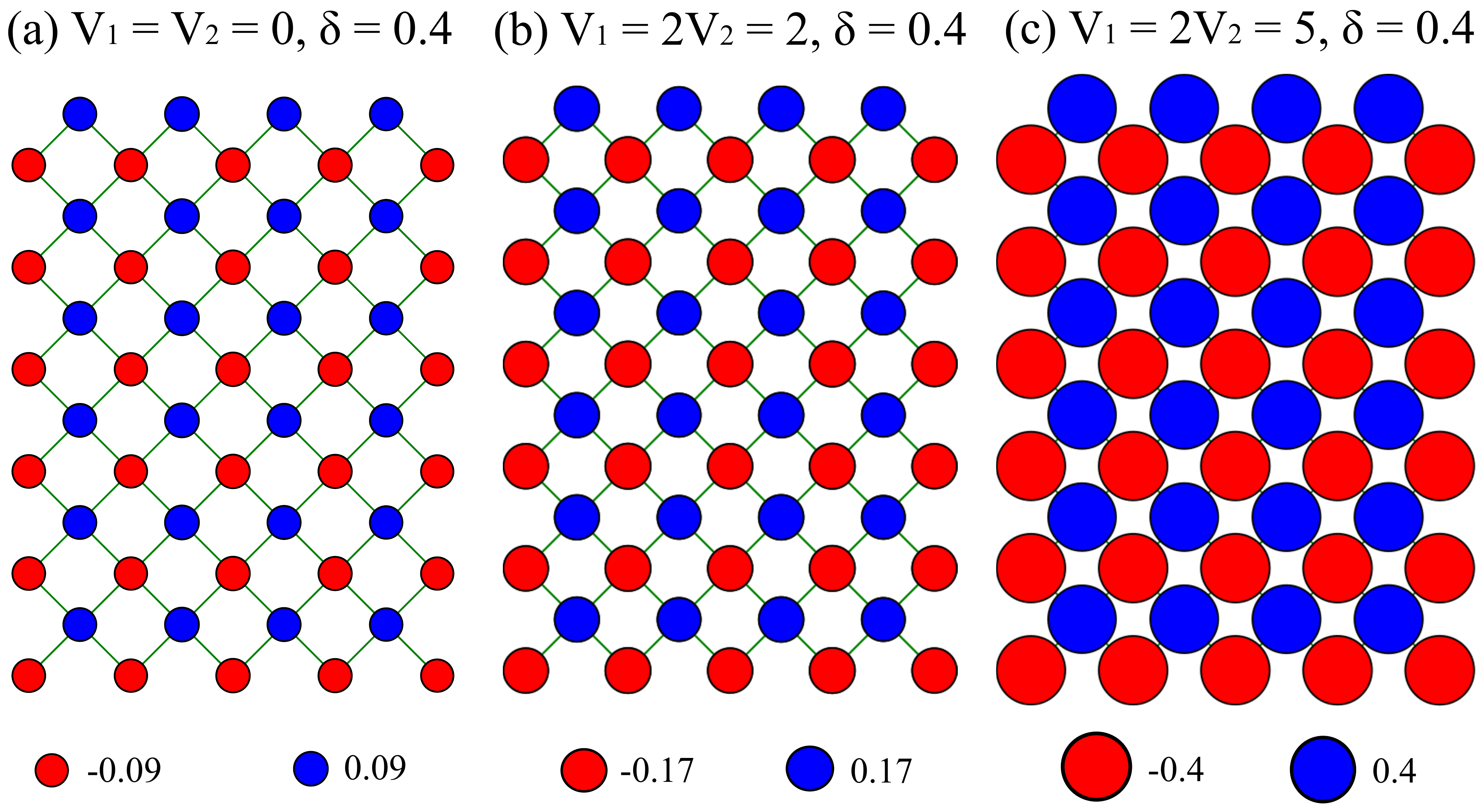}
\caption{Charge density for $\delta = 0.4$ and different repulsive interactions on the $L_y = 6, L_x = 48$ cylinder. We show the results in the bulk of the cylinder for (a) $V_1 = V_2 = 0$, (b) $V_1 = 2V_2 = 2$, and (c) $V_1 = 2V_2 = 5$. All the charge density values have subtracted a constant $0.5$. The negative and positive results are shown as the red and blue colors, respectively. The area of the circle is proportional to the absolute value of the subtracted charge density. At the bottom, the numbers denote the subtracted charge densities.  
}
\label{fig:suppl_density}
\end{figure}

In the main text, we have shown the phase diagram Fig.~\ref{fig:phaseDMRG} with tuning the on-site potential $\delta$ and the repulsive interactions.
Here we show more data for the systems in the larger potential regime.
We choose $\delta = 0.4$ and consider the interactions $V_1 = 2V_2 = 0, 2, 5$.
For $V_1 = V_2 = 0$, the system is a Dirac semimetal. 
With growing interactions, we measure the charge density $\langle n_i \rangle$ and charge hopping energy $\langle c^{\dagger}_i c_j \rangle$.
As shown in Fig.~\ref{fig:suppl_density}, the charge densities of the two sublattices have a small difference $0.18$ in the bulk of the cylinder for $V_1 = V_2 = 0$, due to the on-site potential $\delta = 0.4$.
While for $V_1 = 2V_2 = 2$ the density difference increases slightly to $0.34$, it becomes much larger ($0.8$) at $V_1 = 2V_2 = 5$, which characterizes the emergence of the site-nematic insulating state. 
By computing the charge density, we identify the site-nematic insulating phase in Fig.~\ref{fig:phaseDMRG}.

\begin{figure}[t]
\includegraphics[width = 0.9\linewidth]{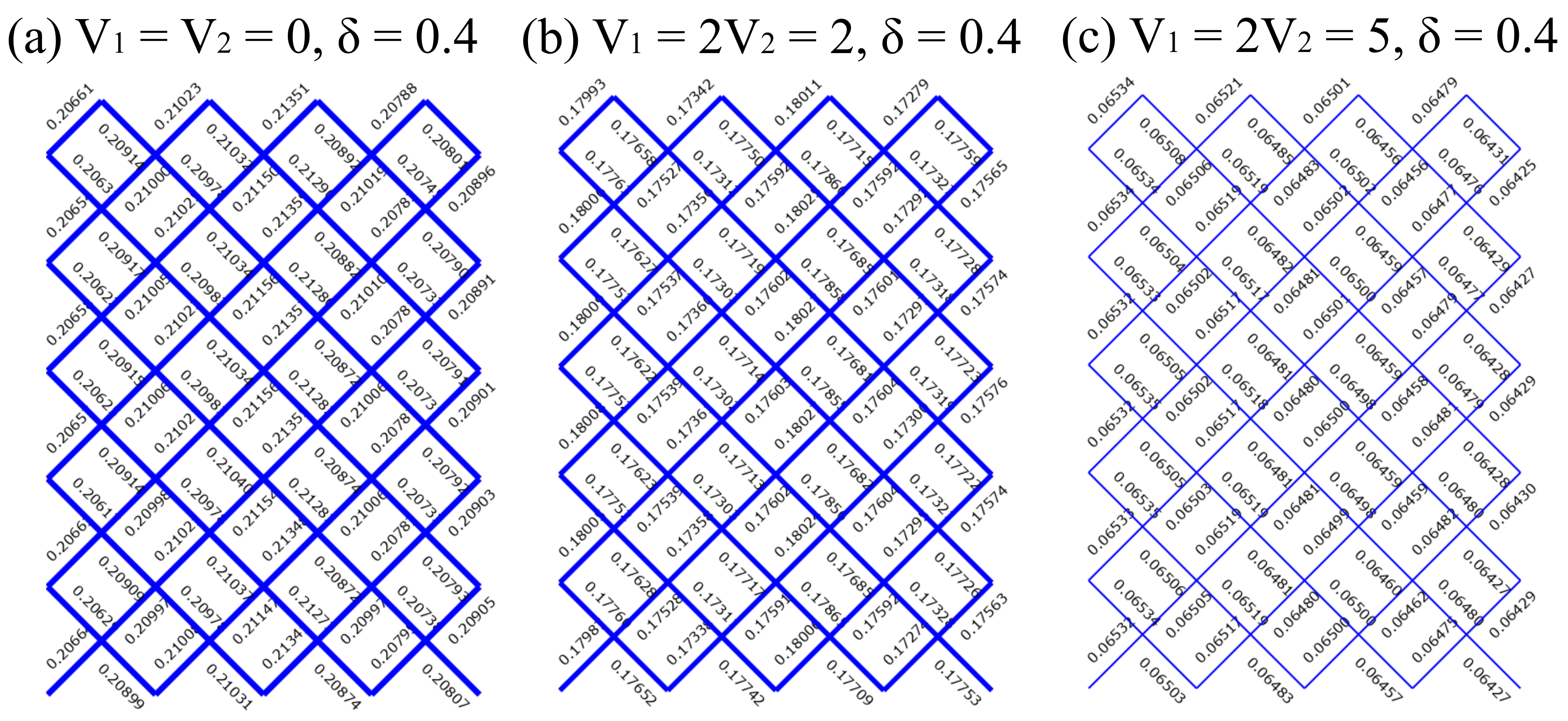}
\caption{Charge hopping energy for $\delta = 0.4$ and different repulsive interactions on the $L_y = 6, L_x = 48$ cylinder. We show the results in the bulk of the cylinder for (a) $V_1 = V_2 = 0$, (b) $V_1 = 2V_2 = 2$, and (c) $V_1 = 2V_2 = 5$. The number denotes the charge hopping energy $\langle c^{\dagger}_i c_j \rangle$ of the corresponding nearest-neighbor bond.  
}
\label{fig:suppl_bond}
\end{figure}

We further compare the nearest-neighbor hopping energy $\langle c^{\dagger}_i c_j \rangle$ in Fig.~\ref{fig:suppl_bond}. For $V_1 = V_2 = 0$ and $V_1 = 2V_2 = 2$, the hopping energies in the bulk seem not to break lattice symmetry and their values are close, which suggests these two parameter points belong to the same phase.
For $V_1 = 2V_2 = 5$, because the charges mainly occupy one of the two sublattices, the nearest-neighbor hopping energy becomes much smaller.

%%%%%%%%%%%%%%%%%%%%%%%%%%%%%%%%%%%%
%%%%%%%%%%%%%%%%%%%%%%%%%%%%%%%%%%%%

%%%%%%%%%%%%%%%%%%%%%%%%%%%%%%%%%%%%
%%%%%%%%%%%%%%%%%%%%%%%%%%%%%%%%%%%%

%%
\section{Anomalous Hall state via  other Dirac semimetal phases} \label{app:other}
\begin{figure}[!t]
\centering
\begin{subfigure}[b]{0.48\textwidth}
\includegraphics[width = \columnwidth]{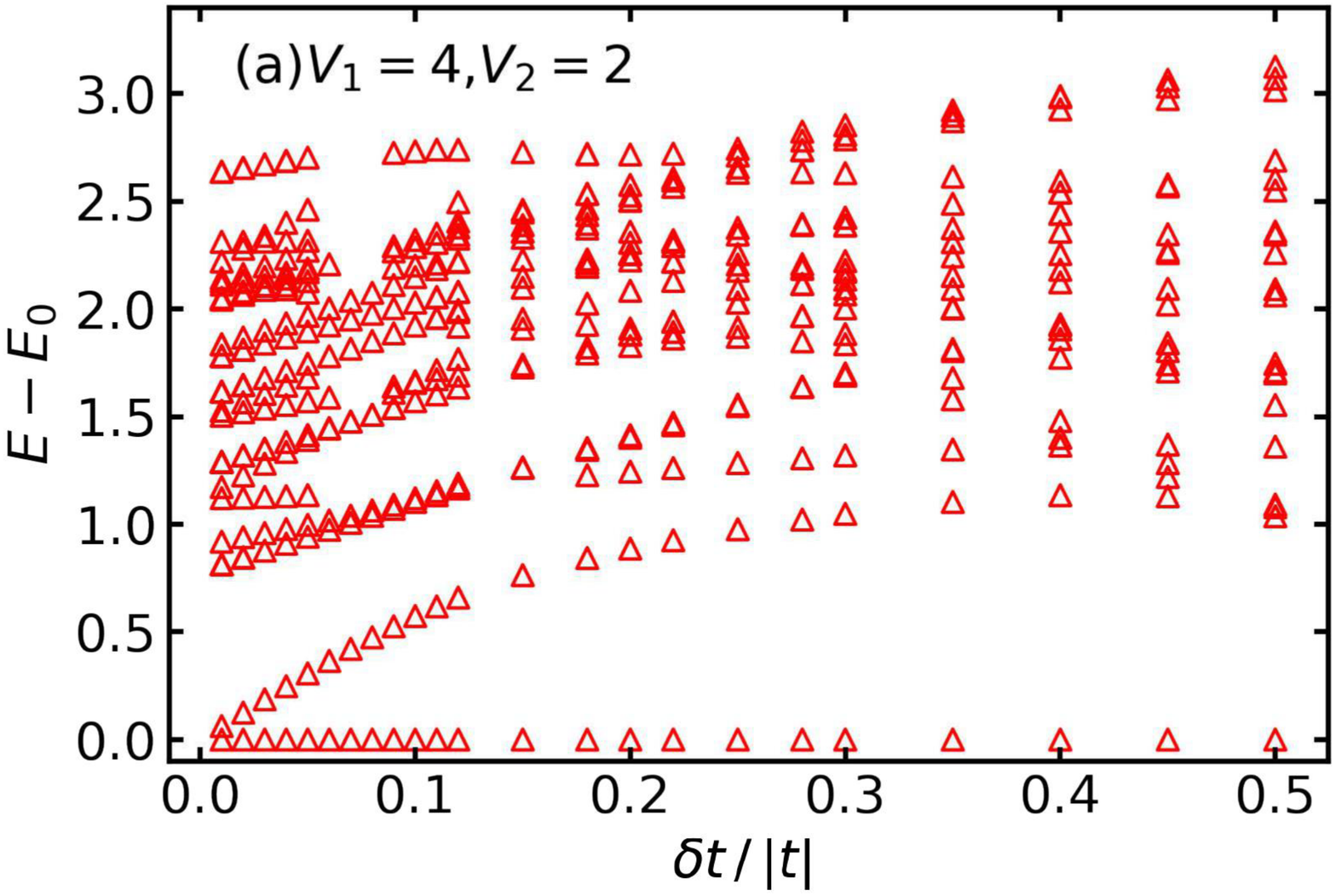}
\caption{}
\end{subfigure}
\hfill
\begin{subfigure}[b]{0.48\textwidth}
\includegraphics[width = \columnwidth]{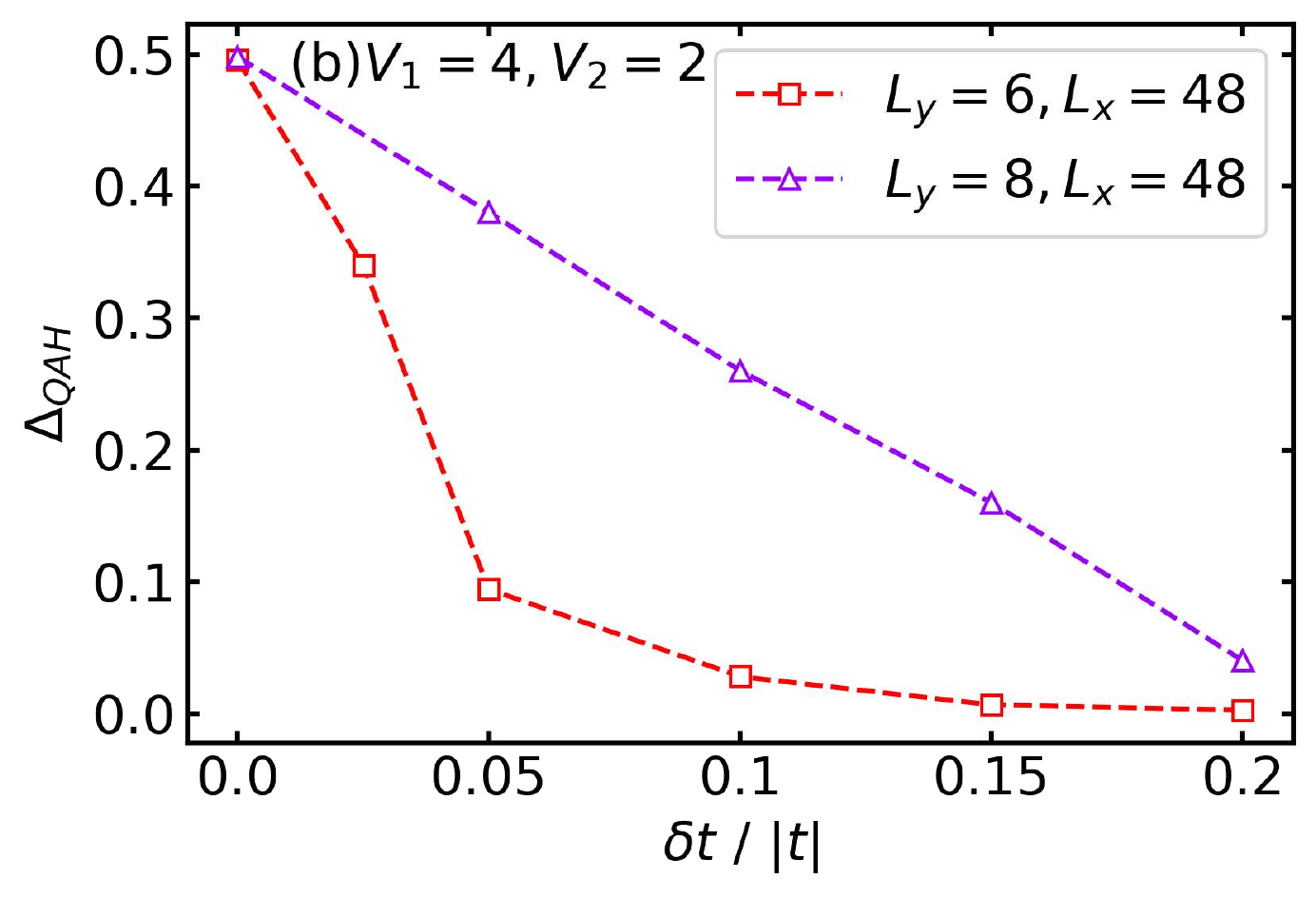}
\caption{}
\end{subfigure}
\hfill
\caption{Robust QAH phase in the presence of bond anisotropy for the model with $V_1 = 4, V_2 = 2$.
(a) Energy spectrum versus the nearest-neighbor bond anisotropy $\delta t$ on the $L_x = L_y = 4$ torus obtained by exact diagonalization calculation. For each parameter point $\delta t / t$, all the energy levels have subtracted its ground-state energy $E_0$.
(b) QAH order parameter $\Delta_{\rm QAH}$ versus $\delta t / t$ obtained by DMRG calculation on the $L_y = 6$ and $L_y = 8$ cylinders, which are obtained using the bond dimension $M = 4000$.}
\label{fig:bond}
\end{figure}

The second anisotropic interaction we consider is the bond anisotropy of the nearest-neighbor (NN) hopping term~\cite{sun2009}.
While we keep the NN hoppings along two directions as $t$ (the diagonal directions with the angles $\pi/4$ and $5\pi/4$ in Fig.~\ref{fig:model}(a)), we set the NN hoppings along the perpendicular directions as $t + \delta t$ (the diagonal directions with the angles $3\pi/4$ and $7\pi/4$).
With a nonzero $\delta t$, the $\mathcal{C}_4$ rotational symmetry of the system reduces to the $\mathcal{C}_2$ symmetry, and the QBT semimetal also becomes a Dirac semimetal with the two Dirac band touching points locating along the diagonal lines in the Brillouin zone~\cite{sun2009}.
Similar to the above case with tuning $\mu$, we start from the QAH phase at $V_1 / t = 4, V_2 / t = 2$ and increase $\delta t$.
The exact diagonalization (ED) energy spectrum and the QAH order parameter obtained by DMRG are shown in Fig.~\ref{fig:bond}.
In the ED spectrum, the gap between the two lowest-energy states and the higher levels decreases with growing $\delta t$.
In DMRG results, the QAH order drops fast with growing $\delta t$ on the $L_y = 6$ system.
On the wider $L_y = 8$ system, $\Delta_{\rm QAH}$ enhances and indicates the strong finite-size effects on $L_y = 6$.
Based on the $L_y = 8$ results, the QAH order can persist in a finite region of $\delta t / t \lesssim 0.2$.
%%%%

%%%
%%%

\clearpage
%%% reference
\bibliography{dirac}

\begin{thebibliography}{49}%
\makeatletter
\providecommand \@ifxundefined [1]{%
 \@ifx{#1\undefined}
}%
\providecommand \@ifnum [1]{%
 \ifnum #1\expandafter \@firstoftwo
 \else \expandafter \@secondoftwo
 \fi
}%
\providecommand \@ifx [1]{%
 \ifx #1\expandafter \@firstoftwo
 \else \expandafter \@secondoftwo
 \fi
}%
\providecommand \natexlab [1]{#1}%
\providecommand \enquote  [1]{``#1''}%
\providecommand \bibnamefont  [1]{#1}%
\providecommand \bibfnamefont [1]{#1}%
\providecommand \citenamefont [1]{#1}%
\providecommand \href@noop [0]{\@secondoftwo}%
\providecommand \href [0]{\begingroup \@sanitize@url \@href}%
\providecommand \@href[1]{\@@startlink{#1}\@@href}%
\providecommand \@@href[1]{\endgroup#1\@@endlink}%
\providecommand \@sanitize@url [0]{\catcode `\\12\catcode `\$12\catcode
  `\&12\catcode `\#12\catcode `\^12\catcode `\_12\catcode `\%12\relax}%
\providecommand \@@startlink[1]{}%
\providecommand \@@endlink[0]{}%
\providecommand \url  [0]{\begingroup\@sanitize@url \@url }%
\providecommand \@url [1]{\endgroup\@href {#1}{\urlprefix }}%
\providecommand \urlprefix  [0]{URL }%
\providecommand \Eprint [0]{\href }%
\providecommand \doibase [0]{http://dx.doi.org/}%
\providecommand \selectlanguage [0]{\@gobble}%
\providecommand \bibinfo  [0]{\@secondoftwo}%
\providecommand \bibfield  [0]{\@secondoftwo}%
\providecommand \translation [1]{[#1]}%
\providecommand \BibitemOpen [0]{}%
\providecommand \bibitemStop [0]{}%
\providecommand \bibitemNoStop [0]{.\EOS\space}%
\providecommand \EOS [0]{\spacefactor3000\relax}%
\providecommand \BibitemShut  [1]{\csname bibitem#1\endcsname}%
\let\auto@bib@innerbib\@empty
%</preamble>
\bibitem [{\citenamefont {Prange}\ and\ \citenamefont
  {Girvin}(2012)}]{prange1990}%
  \BibitemOpen
  \bibfield  {author} {\bibinfo {author} {\bibfnamefont {Richard~E}\
  \bibnamefont {Prange}}\ and\ \bibinfo {author} {\bibfnamefont {Steven~M}\
  \bibnamefont {Girvin}},\ }\href@noop {} {\emph {\bibinfo {title} {The quantum
  Hall effect}}}\ (\bibinfo  {publisher} {Springer Science \& Business Media},\
  \bibinfo {year} {2012})\BibitemShut {NoStop}%
\bibitem [{\citenamefont {Thouless}\ \emph {et~al.}(1982)\citenamefont
  {Thouless}, \citenamefont {Kohmoto}, \citenamefont {Nightingale},\ and\
  \citenamefont {den Nijs}}]{thouless1982}%
  \BibitemOpen
  \bibfield  {author} {\bibinfo {author} {\bibfnamefont {D.~J.}\ \bibnamefont
  {Thouless}}, \bibinfo {author} {\bibfnamefont {M.}~\bibnamefont {Kohmoto}},
  \bibinfo {author} {\bibfnamefont {M.~P.}\ \bibnamefont {Nightingale}}, \ and\
  \bibinfo {author} {\bibfnamefont {M.}~\bibnamefont {den Nijs}},\ }\bibfield
  {title} {\enquote {\bibinfo {title} {Quantized hall conductance in a
  two-dimensional periodic potential},}\ }\href {\doibase
  10.1103/PhysRevLett.49.405} {\bibfield  {journal} {\bibinfo  {journal} {Phys.
  Rev. Lett.}\ }\textbf {\bibinfo {volume} {49}},\ \bibinfo {pages} {405--408}
  (\bibinfo {year} {1982})}\BibitemShut {NoStop}%
\bibitem [{\citenamefont {Haldane}(1988)}]{haldane1988}%
  \BibitemOpen
  \bibfield  {author} {\bibinfo {author} {\bibfnamefont {F.~D.~M.}\
  \bibnamefont {Haldane}},\ }\bibfield  {title} {\enquote {\bibinfo {title}
  {Model for a quantum hall effect without landau levels: Condensed-matter
  realization of the "parity anomaly"},}\ }\href {\doibase
  10.1103/PhysRevLett.61.2015} {\bibfield  {journal} {\bibinfo  {journal}
  {Phys. Rev. Lett.}\ }\textbf {\bibinfo {volume} {61}},\ \bibinfo {pages}
  {2015--2018} (\bibinfo {year} {1988})}\BibitemShut {NoStop}%
\bibitem [{\citenamefont {Raghu}\ \emph {et~al.}(2008)\citenamefont {Raghu},
  \citenamefont {Qi}, \citenamefont {Honerkamp},\ and\ \citenamefont
  {Zhang}}]{raghu2008}%
  \BibitemOpen
  \bibfield  {author} {\bibinfo {author} {\bibfnamefont {S.}~\bibnamefont
  {Raghu}}, \bibinfo {author} {\bibfnamefont {Xiao-Liang}\ \bibnamefont {Qi}},
  \bibinfo {author} {\bibfnamefont {C.}~\bibnamefont {Honerkamp}}, \ and\
  \bibinfo {author} {\bibfnamefont {Shou-Cheng}\ \bibnamefont {Zhang}},\
  }\bibfield  {title} {\enquote {\bibinfo {title} {Topological mott
  insulators},}\ }\href {\doibase 10.1103/PhysRevLett.100.156401} {\bibfield
  {journal} {\bibinfo  {journal} {Phys. Rev. Lett.}\ }\textbf {\bibinfo
  {volume} {100}},\ \bibinfo {pages} {156401} (\bibinfo {year}
  {2008})}\BibitemShut {NoStop}%
\bibitem [{\citenamefont {G{\"o}tz}\ \emph {et~al.}(2018)\citenamefont
  {G{\"o}tz}, \citenamefont {Fijalkowski}, \citenamefont {Pesel}, \citenamefont
  {Hartl}, \citenamefont {Schreyeck}, \citenamefont {Winnerlein}, \citenamefont
  {Grauer}, \citenamefont {Scherer}, \citenamefont {Brunner}, \citenamefont
  {Gould}, \citenamefont {Ahlers},\ and\ \citenamefont {Molenkamp}}]{gotz2018}%
  \BibitemOpen
  \bibfield  {author} {\bibinfo {author} {\bibfnamefont {Martin}\ \bibnamefont
  {G{\"o}tz}}, \bibinfo {author} {\bibfnamefont {Kajetan~M.}\ \bibnamefont
  {Fijalkowski}}, \bibinfo {author} {\bibfnamefont {Eckart}\ \bibnamefont
  {Pesel}}, \bibinfo {author} {\bibfnamefont {Matthias}\ \bibnamefont {Hartl}},
  \bibinfo {author} {\bibfnamefont {Steffen}\ \bibnamefont {Schreyeck}},
  \bibinfo {author} {\bibfnamefont {Martin}\ \bibnamefont {Winnerlein}},
  \bibinfo {author} {\bibfnamefont {Stefan}\ \bibnamefont {Grauer}}, \bibinfo
  {author} {\bibfnamefont {Hansj{\"o}rg}\ \bibnamefont {Scherer}}, \bibinfo
  {author} {\bibfnamefont {Karl}\ \bibnamefont {Brunner}}, \bibinfo {author}
  {\bibfnamefont {Charles}\ \bibnamefont {Gould}}, \bibinfo {author}
  {\bibfnamefont {Franz~J.}\ \bibnamefont {Ahlers}}, \ and\ \bibinfo {author}
  {\bibfnamefont {Laurens~W.}\ \bibnamefont {Molenkamp}},\ }\bibfield  {title}
  {\enquote {\bibinfo {title} {Precision measurement of the quantized anomalous
  hall resistance at zero magnetic field},}\ }\href {\doibase
  10.1063/1.5009718} {\bibfield  {journal} {\bibinfo  {journal} {Applied
  Physics Letters}\ }\textbf {\bibinfo {volume} {112}},\ \bibinfo {pages}
  {072102} (\bibinfo {year} {2018})},\ \Eprint
  {http://arxiv.org/abs/https://doi.org/10.1063/1.5009718}
  {https://doi.org/10.1063/1.5009718} \BibitemShut {NoStop}%
\bibitem [{\citenamefont {Lian}\ \emph {et~al.}(2018)\citenamefont {Lian},
  \citenamefont {Sun}, \citenamefont {Vaezi}, \citenamefont {Qi},\ and\
  \citenamefont {Zhang}}]{lian2018}%
  \BibitemOpen
  \bibfield  {author} {\bibinfo {author} {\bibfnamefont {Biao}\ \bibnamefont
  {Lian}}, \bibinfo {author} {\bibfnamefont {Xiao-Qi}\ \bibnamefont {Sun}},
  \bibinfo {author} {\bibfnamefont {Abolhassan}\ \bibnamefont {Vaezi}},
  \bibinfo {author} {\bibfnamefont {Xiao-Liang}\ \bibnamefont {Qi}}, \ and\
  \bibinfo {author} {\bibfnamefont {Shou-Cheng}\ \bibnamefont {Zhang}},\
  }\bibfield  {title} {\enquote {\bibinfo {title} {Topological quantum
  computation based on chiral majorana fermions},}\ }\href {\doibase
  10.1073/pnas.1810003115} {\bibfield  {journal} {\bibinfo  {journal}
  {Proceedings of the National Academy of Sciences}\ }\textbf {\bibinfo
  {volume} {115}},\ \bibinfo {pages} {10938--10942} (\bibinfo {year} {2018})},\
  \Eprint
  {http://arxiv.org/abs/https://www.pnas.org/content/115/43/10938.full.pdf}
  {https://www.pnas.org/content/115/43/10938.full.pdf} \BibitemShut {NoStop}%
\bibitem [{\citenamefont {Liang}\ \emph {et~al.}(2013)\citenamefont {Liang},
  \citenamefont {Wu},\ and\ \citenamefont {Hu}}]{liang2013}%
  \BibitemOpen
  \bibfield  {author} {\bibinfo {author} {\bibfnamefont {Qi-Feng}\ \bibnamefont
  {Liang}}, \bibinfo {author} {\bibfnamefont {Long-Hua}\ \bibnamefont {Wu}}, \
  and\ \bibinfo {author} {\bibfnamefont {Xiao}\ \bibnamefont {Hu}},\ }\bibfield
   {title} {\enquote {\bibinfo {title} {Electrically tunable topological state
  in [111] perovskite materials with an antiferromagnetic exchange field},}\
  }\href {http://iopscience.iop.org/article/10.1088/1367-2630/15/6/063031/meta}
  {\bibfield  {journal} {\bibinfo  {journal} {New Journal of Physics}\ }\textbf
  {\bibinfo {volume} {15}},\ \bibinfo {pages} {063031} (\bibinfo {year}
  {2013})}\BibitemShut {NoStop}%
\bibitem [{\citenamefont {Yu}\ \emph {et~al.}(2010)\citenamefont {Yu},
  \citenamefont {Zhang}, \citenamefont {Zhang}, \citenamefont {Zhang},
  \citenamefont {Dai},\ and\ \citenamefont {Fang}}]{yu2010}%
  \BibitemOpen
  \bibfield  {author} {\bibinfo {author} {\bibfnamefont {Rui}\ \bibnamefont
  {Yu}}, \bibinfo {author} {\bibfnamefont {Wei}\ \bibnamefont {Zhang}},
  \bibinfo {author} {\bibfnamefont {Hai-Jun}\ \bibnamefont {Zhang}}, \bibinfo
  {author} {\bibfnamefont {Shou-Cheng}\ \bibnamefont {Zhang}}, \bibinfo
  {author} {\bibfnamefont {Xi}~\bibnamefont {Dai}}, \ and\ \bibinfo {author}
  {\bibfnamefont {Zhong}\ \bibnamefont {Fang}},\ }\bibfield  {title} {\enquote
  {\bibinfo {title} {Quantized anomalous hall effect in magnetic topological
  insulators},}\ }\href {\doibase 10.1126/science.1187485} {\bibfield
  {journal} {\bibinfo  {journal} {Science}\ }\textbf {\bibinfo {volume}
  {329}},\ \bibinfo {pages} {61} (\bibinfo {year} {2010})}\BibitemShut
  {NoStop}%
\bibitem [{\citenamefont {{Chang}}\ \emph {et~al.}(2013)\citenamefont
  {{Chang}}, \citenamefont {{Zhang}}, \citenamefont {{Feng}}, \citenamefont
  {{Shen}}, \citenamefont {{Zhang}}, \citenamefont {{Guo}}, \citenamefont
  {{Li}}, \citenamefont {{Ou}}, \citenamefont {{Wei}}, \citenamefont {{Wang}},
  \citenamefont {{Ji}}, \citenamefont {{Feng}}, \citenamefont {{Ji}},
  \citenamefont {{Chen}}, \citenamefont {{Jia}}, \citenamefont {{Dai}},
  \citenamefont {{Fang}}, \citenamefont {{Zhang}}, \citenamefont {{He}},
  \citenamefont {{Wang}}, \citenamefont {{Lu}}, \citenamefont {{Ma}},\ and\
  \citenamefont {{Xue}}}]{chang2013}%
  \BibitemOpen
  \bibfield  {author} {\bibinfo {author} {\bibfnamefont {C.-Z.}\ \bibnamefont
  {{Chang}}}, \bibinfo {author} {\bibfnamefont {J.}~\bibnamefont {{Zhang}}},
  \bibinfo {author} {\bibfnamefont {X.}~\bibnamefont {{Feng}}}, \bibinfo
  {author} {\bibfnamefont {J.}~\bibnamefont {{Shen}}}, \bibinfo {author}
  {\bibfnamefont {Z.}~\bibnamefont {{Zhang}}}, \bibinfo {author} {\bibfnamefont
  {M.}~\bibnamefont {{Guo}}}, \bibinfo {author} {\bibfnamefont
  {K.}~\bibnamefont {{Li}}}, \bibinfo {author} {\bibfnamefont {Y.}~\bibnamefont
  {{Ou}}}, \bibinfo {author} {\bibfnamefont {P.}~\bibnamefont {{Wei}}},
  \bibinfo {author} {\bibfnamefont {L.-L.}\ \bibnamefont {{Wang}}}, \bibinfo
  {author} {\bibfnamefont {Z.-Q.}\ \bibnamefont {{Ji}}}, \bibinfo {author}
  {\bibfnamefont {Y.}~\bibnamefont {{Feng}}}, \bibinfo {author} {\bibfnamefont
  {S.}~\bibnamefont {{Ji}}}, \bibinfo {author} {\bibfnamefont {X.}~\bibnamefont
  {{Chen}}}, \bibinfo {author} {\bibfnamefont {J.}~\bibnamefont {{Jia}}},
  \bibinfo {author} {\bibfnamefont {X.}~\bibnamefont {{Dai}}}, \bibinfo
  {author} {\bibfnamefont {Z.}~\bibnamefont {{Fang}}}, \bibinfo {author}
  {\bibfnamefont {S.-C.}\ \bibnamefont {{Zhang}}}, \bibinfo {author}
  {\bibfnamefont {K.}~\bibnamefont {{He}}}, \bibinfo {author} {\bibfnamefont
  {Y.}~\bibnamefont {{Wang}}}, \bibinfo {author} {\bibfnamefont
  {L.}~\bibnamefont {{Lu}}}, \bibinfo {author} {\bibfnamefont {X.-C.}\
  \bibnamefont {{Ma}}}, \ and\ \bibinfo {author} {\bibfnamefont {Q.-K.}\
  \bibnamefont {{Xue}}},\ }\bibfield  {title} {\enquote {\bibinfo {title}
  {{Experimental Observation of the Quantum Anomalous Hall Effect in a Magnetic
  Topological Insulator}},}\ }\href {\doibase 10.1126/science.1234414}
  {\bibfield  {journal} {\bibinfo  {journal} {Science}\ }\textbf {\bibinfo
  {volume} {340}},\ \bibinfo {pages} {167--170} (\bibinfo {year}
  {2013})}\BibitemShut {NoStop}%
\bibitem [{\citenamefont {Checkelsky}\ \emph {et~al.}(2014)\citenamefont
  {Checkelsky}, \citenamefont {Yoshimi}, \citenamefont {Tsukazaki},
  \citenamefont {Takahashi}, \citenamefont {Kozuka}, \citenamefont {Falson},
  \citenamefont {Kawasaki},\ and\ \citenamefont {Tokura}}]{checkelsky2014}%
  \BibitemOpen
  \bibfield  {author} {\bibinfo {author} {\bibfnamefont {J.~G.}\ \bibnamefont
  {Checkelsky}}, \bibinfo {author} {\bibfnamefont {R.}~\bibnamefont {Yoshimi}},
  \bibinfo {author} {\bibfnamefont {A.}~\bibnamefont {Tsukazaki}}, \bibinfo
  {author} {\bibfnamefont {K.~S.}\ \bibnamefont {Takahashi}}, \bibinfo {author}
  {\bibfnamefont {Y.}~\bibnamefont {Kozuka}}, \bibinfo {author} {\bibfnamefont
  {J.}~\bibnamefont {Falson}}, \bibinfo {author} {\bibfnamefont
  {M.}~\bibnamefont {Kawasaki}}, \ and\ \bibinfo {author} {\bibfnamefont
  {Y.}~\bibnamefont {Tokura}},\ }\bibfield  {title} {\enquote {\bibinfo {title}
  {Trajectory of the anomalous hall effect towards the quantized state in a
  ferromagnetic topological insulator},}\ }\href
  {http://dx.doi.org/10.1038/nphys3053} {\bibfield  {journal} {\bibinfo
  {journal} {Nature Physics}\ }\textbf {\bibinfo {volume} {10}},\ \bibinfo
  {pages} {731} (\bibinfo {year} {2014})}\BibitemShut {NoStop}%
\bibitem [{\citenamefont {Chang}\ \emph {et~al.}(2015)\citenamefont {Chang},
  \citenamefont {Zhao}, \citenamefont {Kim}, \citenamefont {Zhang},
  \citenamefont {Assaf}, \citenamefont {Heiman}, \citenamefont {Zhang},
  \citenamefont {Liu}, \citenamefont {Chan},\ and\ \citenamefont
  {Moodera}}]{chang2015}%
  \BibitemOpen
  \bibfield  {author} {\bibinfo {author} {\bibfnamefont {Cui-Zu}\ \bibnamefont
  {Chang}}, \bibinfo {author} {\bibfnamefont {Weiwei}\ \bibnamefont {Zhao}},
  \bibinfo {author} {\bibfnamefont {Duk~Y.}\ \bibnamefont {Kim}}, \bibinfo
  {author} {\bibfnamefont {Haijun}\ \bibnamefont {Zhang}}, \bibinfo {author}
  {\bibfnamefont {Badih~A.}\ \bibnamefont {Assaf}}, \bibinfo {author}
  {\bibfnamefont {Don}\ \bibnamefont {Heiman}}, \bibinfo {author}
  {\bibfnamefont {Shou-Cheng}\ \bibnamefont {Zhang}}, \bibinfo {author}
  {\bibfnamefont {Chaoxing}\ \bibnamefont {Liu}}, \bibinfo {author}
  {\bibfnamefont {Moses H.~W.}\ \bibnamefont {Chan}}, \ and\ \bibinfo {author}
  {\bibfnamefont {Jagadeesh~S.}\ \bibnamefont {Moodera}},\ }\bibfield  {title}
  {\enquote {\bibinfo {title} {High-precision realization of robust quantum
  anomalous hall state in a hard ferromagnetic topological insulator},}\ }\href
  {http://dx.doi.org/10.1038/nmat4204} {\bibfield  {journal} {\bibinfo
  {journal} {Nature Materials}\ }\textbf {\bibinfo {volume} {14}},\ \bibinfo
  {pages} {473} (\bibinfo {year} {2015})}\BibitemShut {NoStop}%
\bibitem [{\citenamefont {Sun}\ \emph {et~al.}(2009)\citenamefont {Sun},
  \citenamefont {Yao}, \citenamefont {Fradkin},\ and\ \citenamefont
  {Kivelson}}]{sun2009}%
  \BibitemOpen
  \bibfield  {author} {\bibinfo {author} {\bibfnamefont {Kai}\ \bibnamefont
  {Sun}}, \bibinfo {author} {\bibfnamefont {Hong}\ \bibnamefont {Yao}},
  \bibinfo {author} {\bibfnamefont {Eduardo}\ \bibnamefont {Fradkin}}, \ and\
  \bibinfo {author} {\bibfnamefont {Steven~A.}\ \bibnamefont {Kivelson}},\
  }\bibfield  {title} {\enquote {\bibinfo {title} {Topological insulators and
  nematic phases from spontaneous symmetry breaking in 2d fermi systems with a
  quadratic band crossing},}\ }\href {\doibase 10.1103/PhysRevLett.103.046811}
  {\bibfield  {journal} {\bibinfo  {journal} {Phys. Rev. Lett.}\ }\textbf
  {\bibinfo {volume} {103}},\ \bibinfo {pages} {046811} (\bibinfo {year}
  {2009})}\BibitemShut {NoStop}%
\bibitem [{\citenamefont {Nandkishore}\ and\ \citenamefont
  {Levitov}(2010)}]{nandkishore2010}%
  \BibitemOpen
  \bibfield  {author} {\bibinfo {author} {\bibfnamefont {Rahul}\ \bibnamefont
  {Nandkishore}}\ and\ \bibinfo {author} {\bibfnamefont {Leonid}\ \bibnamefont
  {Levitov}},\ }\bibfield  {title} {\enquote {\bibinfo {title} {Quantum
  anomalous hall state in bilayer graphene},}\ }\href {\doibase
  10.1103/PhysRevB.82.115124} {\bibfield  {journal} {\bibinfo  {journal} {Phys.
  Rev. B}\ }\textbf {\bibinfo {volume} {82}},\ \bibinfo {pages} {115124}
  (\bibinfo {year} {2010})}\BibitemShut {NoStop}%
\bibitem [{\citenamefont {Weeks}\ and\ \citenamefont
  {Franz}(2010)}]{weeks2010}%
  \BibitemOpen
  \bibfield  {author} {\bibinfo {author} {\bibfnamefont {C.}~\bibnamefont
  {Weeks}}\ and\ \bibinfo {author} {\bibfnamefont {M.}~\bibnamefont {Franz}},\
  }\bibfield  {title} {\enquote {\bibinfo {title} {Interaction-driven
  instabilities of a dirac semimetal},}\ }\href {\doibase
  10.1103/PhysRevB.81.085105} {\bibfield  {journal} {\bibinfo  {journal} {Phys.
  Rev. B}\ }\textbf {\bibinfo {volume} {81}},\ \bibinfo {pages} {085105}
  (\bibinfo {year} {2010})}\BibitemShut {NoStop}%
\bibitem [{\citenamefont {Wen}\ \emph {et~al.}(2010)\citenamefont {Wen},
  \citenamefont {R\"uegg}, \citenamefont {Wang},\ and\ \citenamefont
  {Fiete}}]{wen2010}%
  \BibitemOpen
  \bibfield  {author} {\bibinfo {author} {\bibfnamefont {Jun}\ \bibnamefont
  {Wen}}, \bibinfo {author} {\bibfnamefont {Andreas}\ \bibnamefont {R\"uegg}},
  \bibinfo {author} {\bibfnamefont {C.-C.~Joseph}\ \bibnamefont {Wang}}, \ and\
  \bibinfo {author} {\bibfnamefont {Gregory~A.}\ \bibnamefont {Fiete}},\
  }\bibfield  {title} {\enquote {\bibinfo {title} {Interaction-driven
  topological insulators on the kagome and the decorated honeycomb lattices},}\
  }\href {\doibase 10.1103/PhysRevB.82.075125} {\bibfield  {journal} {\bibinfo
  {journal} {Phys. Rev. B}\ }\textbf {\bibinfo {volume} {82}},\ \bibinfo
  {pages} {075125} (\bibinfo {year} {2010})}\BibitemShut {NoStop}%
\bibitem [{\citenamefont {Grushin}\ \emph {et~al.}(2013)\citenamefont
  {Grushin}, \citenamefont {Castro}, \citenamefont {Cortijo}, \citenamefont
  {de~Juan}, \citenamefont {Vozmediano},\ and\ \citenamefont
  {Valenzuela}}]{grushin2013}%
  \BibitemOpen
  \bibfield  {author} {\bibinfo {author} {\bibfnamefont {Adolfo~G.}\
  \bibnamefont {Grushin}}, \bibinfo {author} {\bibfnamefont {Eduardo~V.}\
  \bibnamefont {Castro}}, \bibinfo {author} {\bibfnamefont {Alberto}\
  \bibnamefont {Cortijo}}, \bibinfo {author} {\bibfnamefont {Fernando}\
  \bibnamefont {de~Juan}}, \bibinfo {author} {\bibfnamefont {Mar\'{\i}a A.~H.}\
  \bibnamefont {Vozmediano}}, \ and\ \bibinfo {author} {\bibfnamefont
  {Bel\'en}\ \bibnamefont {Valenzuela}},\ }\bibfield  {title} {\enquote
  {\bibinfo {title} {Charge instabilities and topological phases in the
  extended hubbard model on the honeycomb lattice with enlarged unit cell},}\
  }\href {\doibase 10.1103/PhysRevB.87.085136} {\bibfield  {journal} {\bibinfo
  {journal} {Phys. Rev. B}\ }\textbf {\bibinfo {volume} {87}},\ \bibinfo
  {pages} {085136} (\bibinfo {year} {2013})}\BibitemShut {NoStop}%
\bibitem [{\citenamefont {Duri{\'c}}\ \emph {et~al.}(2014)\citenamefont
  {Duri{\'c}}, \citenamefont {Chancellor},\ and\ \citenamefont
  {Herbut}}]{djuric2014}%
  \BibitemOpen
  \bibfield  {author} {\bibinfo {author} {\bibfnamefont {Tanja}\ \bibnamefont
  {Duri{\'c}}}, \bibinfo {author} {\bibfnamefont {Nicholas}\ \bibnamefont
  {Chancellor}}, \ and\ \bibinfo {author} {\bibfnamefont {Igor~F.}\
  \bibnamefont {Herbut}},\ }\bibfield  {title} {\enquote {\bibinfo {title}
  {Interaction-induced anomalous quantum hall state on the honeycomb
  lattice},}\ }\href {\doibase 10.1103/PhysRevB.89.165123} {\bibfield
  {journal} {\bibinfo  {journal} {Phys. Rev. B}\ }\textbf {\bibinfo {volume}
  {89}},\ \bibinfo {pages} {165123} (\bibinfo {year} {2014})}\BibitemShut
  {NoStop}%
\bibitem [{\citenamefont {Garc\'{\i}a-Mart\'{\i}nez}\ \emph
  {et~al.}(2013)\citenamefont {Garc\'{\i}a-Mart\'{\i}nez}, \citenamefont
  {Grushin}, \citenamefont {Neupert}, \citenamefont {Valenzuela},\ and\
  \citenamefont {Castro}}]{garcia2013}%
  \BibitemOpen
  \bibfield  {author} {\bibinfo {author} {\bibfnamefont {Noel~A.}\ \bibnamefont
  {Garc\'{\i}a-Mart\'{\i}nez}}, \bibinfo {author} {\bibfnamefont {Adolfo~G.}\
  \bibnamefont {Grushin}}, \bibinfo {author} {\bibfnamefont {Titus}\
  \bibnamefont {Neupert}}, \bibinfo {author} {\bibfnamefont {Bel\'en}\
  \bibnamefont {Valenzuela}}, \ and\ \bibinfo {author} {\bibfnamefont
  {Eduardo~V.}\ \bibnamefont {Castro}},\ }\bibfield  {title} {\enquote
  {\bibinfo {title} {Interaction-driven phases in the half-filled spinless
  honeycomb lattice from exact diagonalization},}\ }\href {\doibase
  10.1103/PhysRevB.88.245123} {\bibfield  {journal} {\bibinfo  {journal} {Phys.
  Rev. B}\ }\textbf {\bibinfo {volume} {88}},\ \bibinfo {pages} {245123}
  (\bibinfo {year} {2013})}\BibitemShut {NoStop}%
\bibitem [{\citenamefont {Jia}\ \emph {et~al.}(2013)\citenamefont {Jia},
  \citenamefont {Guo}, \citenamefont {Chen}, \citenamefont {Shen},\ and\
  \citenamefont {Feng}}]{jia2013}%
  \BibitemOpen
  \bibfield  {author} {\bibinfo {author} {\bibfnamefont {Yongfei}\ \bibnamefont
  {Jia}}, \bibinfo {author} {\bibfnamefont {Huaiming}\ \bibnamefont {Guo}},
  \bibinfo {author} {\bibfnamefont {Ziyu}\ \bibnamefont {Chen}}, \bibinfo
  {author} {\bibfnamefont {Shun-Qing}\ \bibnamefont {Shen}}, \ and\ \bibinfo
  {author} {\bibfnamefont {Shiping}\ \bibnamefont {Feng}},\ }\bibfield  {title}
  {\enquote {\bibinfo {title} {Effect of interactions on two-dimensional dirac
  fermions},}\ }\href {\doibase 10.1103/PhysRevB.88.075101} {\bibfield
  {journal} {\bibinfo  {journal} {Phys. Rev. B}\ }\textbf {\bibinfo {volume}
  {88}},\ \bibinfo {pages} {075101} (\bibinfo {year} {2013})}\BibitemShut
  {NoStop}%
\bibitem [{\citenamefont {Daghofer}\ and\ \citenamefont
  {Hohenadler}(2014)}]{daghofer2014}%
  \BibitemOpen
  \bibfield  {author} {\bibinfo {author} {\bibfnamefont {Maria}\ \bibnamefont
  {Daghofer}}\ and\ \bibinfo {author} {\bibfnamefont {Martin}\ \bibnamefont
  {Hohenadler}},\ }\bibfield  {title} {\enquote {\bibinfo {title} {Phases of
  correlated spinless fermions on the honeycomb lattice},}\ }\href {\doibase
  10.1103/PhysRevB.89.035103} {\bibfield  {journal} {\bibinfo  {journal} {Phys.
  Rev. B}\ }\textbf {\bibinfo {volume} {89}},\ \bibinfo {pages} {035103}
  (\bibinfo {year} {2014})}\BibitemShut {NoStop}%
\bibitem [{\citenamefont {{Guo}}\ and\ \citenamefont {{Jia}}(2014)}]{guo2014}%
  \BibitemOpen
  \bibfield  {author} {\bibinfo {author} {\bibfnamefont {H.}~\bibnamefont
  {{Guo}}}\ and\ \bibinfo {author} {\bibfnamefont {Y.}~\bibnamefont {{Jia}}},\
  }\bibfield  {title} {\enquote {\bibinfo {title} {{Interaction-driven phases
  in a Dirac semimetal: exact diagonalization results}},}\ }\href {\doibase
  10.1088/0953-8984/26/47/475601} {\bibfield  {journal} {\bibinfo  {journal}
  {Journal of Physics Condensed Matter}\ }\textbf {\bibinfo {volume} {26}},\
  \bibinfo {eid} {475601} (\bibinfo {year} {2014})},\ \Eprint
  {http://arxiv.org/abs/1402.4274} {arXiv:1402.4274 [cond-mat.str-el]}
  \BibitemShut {NoStop}%
\bibitem [{\citenamefont {Motruk}\ \emph {et~al.}(2015)\citenamefont {Motruk},
  \citenamefont {Grushin}, \citenamefont {de~Juan},\ and\ \citenamefont
  {Pollmann}}]{motruk2015}%
  \BibitemOpen
  \bibfield  {author} {\bibinfo {author} {\bibfnamefont {Johannes}\
  \bibnamefont {Motruk}}, \bibinfo {author} {\bibfnamefont {Adolfo~G.}\
  \bibnamefont {Grushin}}, \bibinfo {author} {\bibfnamefont {Fernando}\
  \bibnamefont {de~Juan}}, \ and\ \bibinfo {author} {\bibfnamefont {Frank}\
  \bibnamefont {Pollmann}},\ }\bibfield  {title} {\enquote {\bibinfo {title}
  {Interaction-driven phases in the half-filled honeycomb lattice: An infinite
  density matrix renormalization group study},}\ }\href {\doibase
  10.1103/PhysRevB.92.085147} {\bibfield  {journal} {\bibinfo  {journal} {Phys.
  Rev. B}\ }\textbf {\bibinfo {volume} {92}},\ \bibinfo {pages} {085147}
  (\bibinfo {year} {2015})}\BibitemShut {NoStop}%
\bibitem [{\citenamefont {Capponi}\ and\ \citenamefont
  {L\"auchli}(2015)}]{capponi2015}%
  \BibitemOpen
  \bibfield  {author} {\bibinfo {author} {\bibfnamefont {Sylvain}\ \bibnamefont
  {Capponi}}\ and\ \bibinfo {author} {\bibfnamefont {Andreas~M.}\ \bibnamefont
  {L\"auchli}},\ }\bibfield  {title} {\enquote {\bibinfo {title} {Phase diagram
  of interacting spinless fermions on the honeycomb lattice: A comprehensive
  exact diagonalization study},}\ }\href {\doibase 10.1103/PhysRevB.92.085146}
  {\bibfield  {journal} {\bibinfo  {journal} {Phys. Rev. B}\ }\textbf {\bibinfo
  {volume} {92}},\ \bibinfo {pages} {085146} (\bibinfo {year}
  {2015})}\BibitemShut {NoStop}%
\bibitem [{\citenamefont {Scherer}\ \emph {et~al.}(2015)\citenamefont
  {Scherer}, \citenamefont {Scherer},\ and\ \citenamefont
  {Honerkamp}}]{scherer2015}%
  \BibitemOpen
  \bibfield  {author} {\bibinfo {author} {\bibfnamefont {Daniel~D.}\
  \bibnamefont {Scherer}}, \bibinfo {author} {\bibfnamefont {Michael~M.}\
  \bibnamefont {Scherer}}, \ and\ \bibinfo {author} {\bibfnamefont {Carsten}\
  \bibnamefont {Honerkamp}},\ }\bibfield  {title} {\enquote {\bibinfo {title}
  {Correlated spinless fermions on the honeycomb lattice revisited},}\ }\href
  {\doibase 10.1103/PhysRevB.92.155137} {\bibfield  {journal} {\bibinfo
  {journal} {Phys. Rev. B}\ }\textbf {\bibinfo {volume} {92}},\ \bibinfo
  {pages} {155137} (\bibinfo {year} {2015})}\BibitemShut {NoStop}%
\bibitem [{\citenamefont {Chong}\ \emph {et~al.}(2008)\citenamefont {Chong},
  \citenamefont {Wen},\ and\ \citenamefont {Solja\ifmmode \check{c}\else
  \v{c}\fi{}i\ifmmode~\acute{c}\else \'{c}\fi{}}}]{chong2008}%
  \BibitemOpen
  \bibfield  {author} {\bibinfo {author} {\bibfnamefont {Y.~D.}\ \bibnamefont
  {Chong}}, \bibinfo {author} {\bibfnamefont {Xiao-Gang}\ \bibnamefont {Wen}},
  \ and\ \bibinfo {author} {\bibfnamefont {Marin}\ \bibnamefont {Solja\ifmmode
  \check{c}\else \v{c}\fi{}i\ifmmode~\acute{c}\else \'{c}\fi{}}},\ }\bibfield
  {title} {\enquote {\bibinfo {title} {Effective theory of quadratic
  degeneracies},}\ }\href {\doibase 10.1103/PhysRevB.77.235125} {\bibfield
  {journal} {\bibinfo  {journal} {Phys. Rev. B}\ }\textbf {\bibinfo {volume}
  {77}},\ \bibinfo {pages} {235125} (\bibinfo {year} {2008})}\BibitemShut
  {NoStop}%
\bibitem [{\citenamefont {Sun}\ and\ \citenamefont {Fradkin}(2008)}]{sun2008}%
  \BibitemOpen
  \bibfield  {author} {\bibinfo {author} {\bibfnamefont {Kai}\ \bibnamefont
  {Sun}}\ and\ \bibinfo {author} {\bibfnamefont {Eduardo}\ \bibnamefont
  {Fradkin}},\ }\bibfield  {title} {\enquote {\bibinfo {title} {Time-reversal
  symmetry breaking and spontaneous anomalous hall effect in fermi fluids},}\
  }\href {\doibase 10.1103/PhysRevB.78.245122} {\bibfield  {journal} {\bibinfo
  {journal} {Phys. Rev. B}\ }\textbf {\bibinfo {volume} {78}},\ \bibinfo
  {pages} {245122} (\bibinfo {year} {2008})}\BibitemShut {NoStop}%
\bibitem [{\citenamefont {Tsai}\ \emph {et~al.}(2015)\citenamefont {Tsai},
  \citenamefont {Fang}, \citenamefont {Yao},\ and\ \citenamefont
  {Hu}}]{tsai2015}%
  \BibitemOpen
  \bibfield  {author} {\bibinfo {author} {\bibfnamefont {Wei-Feng}\
  \bibnamefont {Tsai}}, \bibinfo {author} {\bibfnamefont {Chen}\ \bibnamefont
  {Fang}}, \bibinfo {author} {\bibfnamefont {Hong}\ \bibnamefont {Yao}}, \ and\
  \bibinfo {author} {\bibfnamefont {Jiangping}\ \bibnamefont {Hu}},\ }\bibfield
   {title} {\enquote {\bibinfo {title} {Interaction-driven topological and
  nematic phases on the lieb lattice},}\ }\href
  {http://iopscience.iop.org/article/10.1088/1367-2630/17/5/055016/meta}
  {\bibfield  {journal} {\bibinfo  {journal} {New Journal of Physics}\ }\textbf
  {\bibinfo {volume} {17}},\ \bibinfo {pages} {055016} (\bibinfo {year}
  {2015})}\BibitemShut {NoStop}%
\bibitem [{\citenamefont {Wu}\ \emph {et~al.}(2016)\citenamefont {Wu},
  \citenamefont {He}, \citenamefont {Fang}, \citenamefont {Meng},\ and\
  \citenamefont {Lu}}]{wu2016}%
  \BibitemOpen
  \bibfield  {author} {\bibinfo {author} {\bibfnamefont {Han-Qing}\
  \bibnamefont {Wu}}, \bibinfo {author} {\bibfnamefont {Yuan-Yao}\ \bibnamefont
  {He}}, \bibinfo {author} {\bibfnamefont {Chen}\ \bibnamefont {Fang}},
  \bibinfo {author} {\bibfnamefont {Zi~Yang}\ \bibnamefont {Meng}}, \ and\
  \bibinfo {author} {\bibfnamefont {Zhong-Yi}\ \bibnamefont {Lu}},\ }\bibfield
  {title} {\enquote {\bibinfo {title} {Diagnosis of interaction-driven
  topological phase via exact diagonalization},}\ }\href {\doibase
  10.1103/PhysRevLett.117.066403} {\bibfield  {journal} {\bibinfo  {journal}
  {Phys. Rev. Lett.}\ }\textbf {\bibinfo {volume} {117}},\ \bibinfo {pages}
  {066403} (\bibinfo {year} {2016})}\BibitemShut {NoStop}%
\bibitem [{\citenamefont {Zhu}\ \emph {et~al.}(2016)\citenamefont {Zhu},
  \citenamefont {Gong}, \citenamefont {Zeng}, \citenamefont {Fu},\ and\
  \citenamefont {Sheng}}]{zhu2016}%
  \BibitemOpen
  \bibfield  {author} {\bibinfo {author} {\bibfnamefont {W.}~\bibnamefont
  {Zhu}}, \bibinfo {author} {\bibfnamefont {Shou-Shu}\ \bibnamefont {Gong}},
  \bibinfo {author} {\bibfnamefont {Tian-Sheng}\ \bibnamefont {Zeng}}, \bibinfo
  {author} {\bibfnamefont {Liang}\ \bibnamefont {Fu}}, \ and\ \bibinfo {author}
  {\bibfnamefont {D.~N.}\ \bibnamefont {Sheng}},\ }\bibfield  {title} {\enquote
  {\bibinfo {title} {Interaction-driven spontaneous quantum hall effect on a
  kagome lattice},}\ }\href {\doibase 10.1103/PhysRevLett.117.096402}
  {\bibfield  {journal} {\bibinfo  {journal} {Phys. Rev. Lett.}\ }\textbf
  {\bibinfo {volume} {117}},\ \bibinfo {pages} {096402} (\bibinfo {year}
  {2016})}\BibitemShut {NoStop}%
\bibitem [{\citenamefont {Sur}\ \emph {et~al.}(2018)\citenamefont {Sur},
  \citenamefont {Gong}, \citenamefont {Yang},\ and\ \citenamefont
  {Vafek}}]{sur2018}%
  \BibitemOpen
  \bibfield  {author} {\bibinfo {author} {\bibfnamefont {Shouvik}\ \bibnamefont
  {Sur}}, \bibinfo {author} {\bibfnamefont {Shou-Shu}\ \bibnamefont {Gong}},
  \bibinfo {author} {\bibfnamefont {Kun}\ \bibnamefont {Yang}}, \ and\ \bibinfo
  {author} {\bibfnamefont {Oskar}\ \bibnamefont {Vafek}},\ }\bibfield  {title}
  {\enquote {\bibinfo {title} {Quantum anomalous hall insulator stabilized by
  competing interactions},}\ }\href {\doibase 10.1103/PhysRevB.98.125144}
  {\bibfield  {journal} {\bibinfo  {journal} {Phys. Rev. B}\ }\textbf {\bibinfo
  {volume} {98}},\ \bibinfo {pages} {125144} (\bibinfo {year}
  {2018})}\BibitemShut {NoStop}%
\bibitem [{\citenamefont {Zeng}\ \emph {et~al.}(2018)\citenamefont {Zeng},
  \citenamefont {Zhu},\ and\ \citenamefont {Sheng}}]{zeng2018}%
  \BibitemOpen
  \bibfield  {author} {\bibinfo {author} {\bibfnamefont {Tian-Sheng}\
  \bibnamefont {Zeng}}, \bibinfo {author} {\bibfnamefont {W.}~\bibnamefont
  {Zhu}}, \ and\ \bibinfo {author} {\bibfnamefont {D.~N.}\ \bibnamefont
  {Sheng}},\ }\bibfield  {title} {\enquote {\bibinfo {title} {Tuning
  topological phase and quantum anomalous hall effect by interaction in
  quadratic band touching systems},}\ }\href
  {http://www.nature.com/articles/s41535-018-0120-5} {\bibfield  {journal}
  {\bibinfo  {journal} {npj Quantum Materials}\ }\textbf {\bibinfo {volume}
  {3}},\ \bibinfo {pages} {49} (\bibinfo {year} {2018})}\BibitemShut {NoStop}%
\bibitem [{\citenamefont {Liang}\ \emph {et~al.}(2017)\citenamefont {Liang},
  \citenamefont {Zhou}, \citenamefont {Yu}, \citenamefont {Wang},\ and\
  \citenamefont {Weng}}]{liang2017}%
  \BibitemOpen
  \bibfield  {author} {\bibinfo {author} {\bibfnamefont {Qi-Feng}\ \bibnamefont
  {Liang}}, \bibinfo {author} {\bibfnamefont {Jian}\ \bibnamefont {Zhou}},
  \bibinfo {author} {\bibfnamefont {Rui}\ \bibnamefont {Yu}}, \bibinfo {author}
  {\bibfnamefont {Xi}~\bibnamefont {Wang}}, \ and\ \bibinfo {author}
  {\bibfnamefont {Hongming}\ \bibnamefont {Weng}},\ }\bibfield  {title}
  {\enquote {\bibinfo {title} {Interaction-driven quantum anomalous hall effect
  in halogenated hematite nanosheets},}\ }\href {\doibase
  10.1103/PhysRevB.96.205412} {\bibfield  {journal} {\bibinfo  {journal} {Phys.
  Rev. B}\ }\textbf {\bibinfo {volume} {96}},\ \bibinfo {pages} {205412}
  (\bibinfo {year} {2017})}\BibitemShut {NoStop}%
\bibitem [{\citenamefont {Wang}\ \emph {et~al.}(2018)\citenamefont {Wang},
  \citenamefont {Liu}, \citenamefont {Yang},\ and\ \citenamefont
  {Liu}}]{wang2018}%
  \BibitemOpen
  \bibfield  {author} {\bibinfo {author} {\bibfnamefont {ZF}~\bibnamefont
  {Wang}}, \bibinfo {author} {\bibfnamefont {Zhao}\ \bibnamefont {Liu}},
  \bibinfo {author} {\bibfnamefont {Jinlong}\ \bibnamefont {Yang}}, \ and\
  \bibinfo {author} {\bibfnamefont {Feng}\ \bibnamefont {Liu}},\ }\bibfield
  {title} {\enquote {\bibinfo {title} {Light-induced type-ii band inversion and
  quantum anomalous hall state in monolayer fese},}\ }\href {\doibase
  10.1103/PhysRevLett.120.156406} {\bibfield  {journal} {\bibinfo  {journal}
  {Phys. Rev. Lett.}\ }\textbf {\bibinfo {volume} {120}},\ \bibinfo {pages}
  {156406} (\bibinfo {year} {2018})}\BibitemShut {NoStop}%
\bibitem [{\citenamefont {Osada}(2019)}]{osada2019}%
  \BibitemOpen
  \bibfield  {author} {\bibinfo {author} {\bibfnamefont {Toshihito}\
  \bibnamefont {Osada}},\ }\bibfield  {title} {\enquote {\bibinfo {title}
  {Topological properties of $\tau$-type organic conductors with a checkerboard
  lattice},}\ }\href {\doibase 10.7566/JPSJ.88.114707} {\bibfield  {journal}
  {\bibinfo  {journal} {J. Phys. Soc. Jpn.}\ }\textbf {\bibinfo {volume}
  {88}},\ \bibinfo {pages} {114707} (\bibinfo {year} {2019})}\BibitemShut
  {NoStop}%
\bibitem [{\citenamefont {Wirth}\ \emph {et~al.}(2011)\citenamefont {Wirth},
  \citenamefont {{\"O}lschl{\"a}ger},\ and\ \citenamefont
  {Hemmerich}}]{wirth2011}%
  \BibitemOpen
  \bibfield  {author} {\bibinfo {author} {\bibfnamefont {Georg}\ \bibnamefont
  {Wirth}}, \bibinfo {author} {\bibfnamefont {Matthias}\ \bibnamefont
  {{\"O}lschl{\"a}ger}}, \ and\ \bibinfo {author} {\bibfnamefont {Andreas}\
  \bibnamefont {Hemmerich}},\ }\bibfield  {title} {\enquote {\bibinfo {title}
  {Evidence for orbital superfluidity in the p-band of a bipartite optical
  square lattice},}\ }\href {\doibase 10.1038/nphys1857} {\bibfield  {journal}
  {\bibinfo  {journal} {Nature Physics}\ }\textbf {\bibinfo {volume} {7}},\
  \bibinfo {pages} {147--153} (\bibinfo {year} {2011})}\BibitemShut {NoStop}%
\bibitem [{\citenamefont {Mielke}(1991)}]{mielke1991}%
  \BibitemOpen
  \bibfield  {author} {\bibinfo {author} {\bibfnamefont {A}~\bibnamefont
  {Mielke}},\ }\bibfield  {title} {\enquote {\bibinfo {title} {Ferromagnetism
  in the hubbard model on line graphs and further considerations},}\ }\href
  {\doibase 10.1088/0305-4470/24/14/018} {\bibfield  {journal} {\bibinfo
  {journal} {Journal of Physics A: Mathematical and General}\ }\textbf
  {\bibinfo {volume} {24}},\ \bibinfo {pages} {3311} (\bibinfo {year}
  {1991})}\BibitemShut {NoStop}%
\bibitem [{\citenamefont {Montambaux}\ \emph {et~al.}(2018)\citenamefont
  {Montambaux}, \citenamefont {Lim}, \citenamefont {Fuchs},\ and\ \citenamefont
  {Pi{\'e}chon}}]{montambaux2018}%
  \BibitemOpen
  \bibfield  {author} {\bibinfo {author} {\bibfnamefont {Gilles}\ \bibnamefont
  {Montambaux}}, \bibinfo {author} {\bibfnamefont {Lih-King}\ \bibnamefont
  {Lim}}, \bibinfo {author} {\bibfnamefont {Jean-No{\"e}l}\ \bibnamefont
  {Fuchs}}, \ and\ \bibinfo {author} {\bibfnamefont {Fr{\'e}d{\'e}ric}\
  \bibnamefont {Pi{\'e}chon}},\ }\bibfield  {title} {\enquote {\bibinfo {title}
  {Winding vector: how to annihilate two dirac points with the same charge},}\
  }\href {\doibase 10.1103/PhysRevLett.121.256402} {\bibfield  {journal}
  {\bibinfo  {journal} {Physical review letters}\ }\textbf {\bibinfo {volume}
  {121}},\ \bibinfo {pages} {256402} (\bibinfo {year} {2018})}\BibitemShut
  {NoStop}%
\bibitem [{\citenamefont {Iskin}(2019)}]{iskin2019}%
  \BibitemOpen
  \bibfield  {author} {\bibinfo {author} {\bibfnamefont {M}~\bibnamefont
  {Iskin}},\ }\bibfield  {title} {\enquote {\bibinfo {title} {Origin of
  flat-band superfluidity on the mielke checkerboard lattice},}\ }\href
  {\doibase 10.1103/PhysRevA.99.053608} {\bibfield  {journal} {\bibinfo
  {journal} {Physical Review A}\ }\textbf {\bibinfo {volume} {99}},\ \bibinfo
  {pages} {053608} (\bibinfo {year} {2019})}\BibitemShut {NoStop}%
\bibitem [{\citenamefont {Lim}\ \emph {et~al.}(2012)\citenamefont {Lim},
  \citenamefont {Fuchs},\ and\ \citenamefont {Montambaux}}]{lim2012}%
  \BibitemOpen
  \bibfield  {author} {\bibinfo {author} {\bibfnamefont {Lih-King}\
  \bibnamefont {Lim}}, \bibinfo {author} {\bibfnamefont {Jean-No{\"e}l}\
  \bibnamefont {Fuchs}}, \ and\ \bibinfo {author} {\bibfnamefont {Gilles}\
  \bibnamefont {Montambaux}},\ }\bibfield  {title} {\enquote {\bibinfo {title}
  {Bloch-zener oscillations across a merging transition of dirac points},}\
  }\href {\doibase 10.1103/PhysRevLett.108.175303} {\bibfield  {journal}
  {\bibinfo  {journal} {Phys. Rev. Lett.}\ }\textbf {\bibinfo {volume} {108}},\
  \bibinfo {pages} {175303} (\bibinfo {year} {2012})}\BibitemShut {NoStop}%
\bibitem [{\citenamefont {Isobe}\ \emph {et~al.}(2016)\citenamefont {Isobe},
  \citenamefont {Yang}, \citenamefont {Chubukov}, \citenamefont {Schmalian},\
  and\ \citenamefont {Nagaosa}}]{isobe2016}%
  \BibitemOpen
  \bibfield  {author} {\bibinfo {author} {\bibfnamefont {Hiroki}\ \bibnamefont
  {Isobe}}, \bibinfo {author} {\bibfnamefont {Bohm-Jung}\ \bibnamefont {Yang}},
  \bibinfo {author} {\bibfnamefont {Andrey}\ \bibnamefont {Chubukov}}, \bibinfo
  {author} {\bibfnamefont {J{\"o}rg}\ \bibnamefont {Schmalian}}, \ and\
  \bibinfo {author} {\bibfnamefont {Naoto}\ \bibnamefont {Nagaosa}},\
  }\bibfield  {title} {\enquote {\bibinfo {title} {Emergent non-fermi-liquid at
  the quantum critical point of a topological phase transition in two
  dimensions},}\ }\href {\doibase 10.1103/PhysRevLett.116.076803} {\bibfield
  {journal} {\bibinfo  {journal} {Phys. Rev. Lett.}\ }\textbf {\bibinfo
  {volume} {116}},\ \bibinfo {pages} {076803} (\bibinfo {year}
  {2016})}\BibitemShut {NoStop}%
\bibitem [{\citenamefont {Sur}\ and\ \citenamefont {Roy}(2019)}]{sur2019}%
  \BibitemOpen
  \bibfield  {author} {\bibinfo {author} {\bibfnamefont {Shouvik}\ \bibnamefont
  {Sur}}\ and\ \bibinfo {author} {\bibfnamefont {Bitan}\ \bibnamefont {Roy}},\
  }\bibfield  {title} {\enquote {\bibinfo {title} {Unifying interacting nodal
  semimetals: A new route to strong coupling},}\ }\href {\doibase
  10.1103/PhysRevLett.123.207601} {\bibfield  {journal} {\bibinfo  {journal}
  {Phys. Rev. Lett.}\ }\textbf {\bibinfo {volume} {123}},\ \bibinfo {pages}
  {207601} (\bibinfo {year} {2019})}\BibitemShut {NoStop}%
\bibitem [{\citenamefont {Sachdev}(2011)}]{sachdevBook}%
  \BibitemOpen
  \bibfield  {author} {\bibinfo {author} {\bibfnamefont {Subir}\ \bibnamefont
  {Sachdev}},\ }\href {\doibase 10.1017/CBO9780511973765} {\emph {\bibinfo
  {title} {Quantum phase transitions}}}\ (\bibinfo  {publisher} {Cambridge
  university press},\ \bibinfo {year} {2011})\BibitemShut {NoStop}%
\bibitem [{\citenamefont {White}(1992)}]{white1992}%
  \BibitemOpen
  \bibfield  {author} {\bibinfo {author} {\bibfnamefont {Steven~R.}\
  \bibnamefont {White}},\ }\bibfield  {title} {\enquote {\bibinfo {title}
  {Density matrix formulation for quantum renormalization groups},}\ }\href
  {\doibase 10.1103/PhysRevLett.69.2863} {\bibfield  {journal} {\bibinfo
  {journal} {Phys. Rev. Lett.}\ }\textbf {\bibinfo {volume} {69}},\ \bibinfo
  {pages} {2863--2866} (\bibinfo {year} {1992})}\BibitemShut {NoStop}%
\bibitem [{\citenamefont {Jiang}\ \emph {et~al.}(2012)\citenamefont {Jiang},
  \citenamefont {Wang},\ and\ \citenamefont {Balents}}]{jiang2012}%
  \BibitemOpen
  \bibfield  {author} {\bibinfo {author} {\bibfnamefont {Hong-Chen}\
  \bibnamefont {Jiang}}, \bibinfo {author} {\bibfnamefont {Zhenghan}\
  \bibnamefont {Wang}}, \ and\ \bibinfo {author} {\bibfnamefont {Leon}\
  \bibnamefont {Balents}},\ }\bibfield  {title} {\enquote {\bibinfo {title}
  {Identifying topological order by entanglement entropy},}\ }\href {\doibase
  10.1038/nphys2465} {\bibfield  {journal} {\bibinfo  {journal} {Nature
  Physics}\ }\textbf {\bibinfo {volume} {8}},\ \bibinfo {pages} {902--905}
  (\bibinfo {year} {2012})}\BibitemShut {NoStop}%
\bibitem [{\citenamefont {Gong}\ \emph {et~al.}(2014)\citenamefont {Gong},
  \citenamefont {Zhu},\ and\ \citenamefont {Sheng}}]{gong2014kagome}%
  \BibitemOpen
  \bibfield  {author} {\bibinfo {author} {\bibfnamefont {Shou-Shu}\
  \bibnamefont {Gong}}, \bibinfo {author} {\bibfnamefont {Wei}\ \bibnamefont
  {Zhu}}, \ and\ \bibinfo {author} {\bibfnamefont {DN}~\bibnamefont {Sheng}},\
  }\bibfield  {title} {\enquote {\bibinfo {title} {Emergent chiral spin liquid:
  Fractional quantum hall effect in a kagome heisenberg model},}\ }\href
  {https://www.nature.com/articles/srep06317} {\bibfield  {journal} {\bibinfo
  {journal} {Scientific reports}\ }\textbf {\bibinfo {volume} {4}},\ \bibinfo
  {pages} {6317} (\bibinfo {year} {2014})}\BibitemShut {NoStop}%
\bibitem [{\citenamefont {Zaletel}\ \emph {et~al.}(2014)\citenamefont
  {Zaletel}, \citenamefont {Mong},\ and\ \citenamefont
  {Pollmann}}]{zaletel2014}%
  \BibitemOpen
  \bibfield  {author} {\bibinfo {author} {\bibfnamefont {Michael}\ \bibnamefont
  {Zaletel}}, \bibinfo {author} {\bibfnamefont {Roger}\ \bibnamefont {Mong}}, \
  and\ \bibinfo {author} {\bibfnamefont {Frank}\ \bibnamefont {Pollmann}},\
  }\bibfield  {title} {\enquote {\bibinfo {title} {Flux insertion,
  entanglement, and quantized responses},}\ }\href
  {http://iopscience.iop.org/article/10.1088/1742-5468/2014/10/P10007/meta}
  {\bibfield  {journal} {\bibinfo  {journal} {Journal of Statistical Mechanics:
  Theory and Experiment}\ }\textbf {\bibinfo {volume} {2014}},\ \bibinfo
  {pages} {P10007} (\bibinfo {year} {2014})}\BibitemShut {NoStop}%
\bibitem [{\citenamefont {Laughlin}(1981)}]{laughlin1981}%
  \BibitemOpen
  \bibfield  {author} {\bibinfo {author} {\bibfnamefont {R.~B.}\ \bibnamefont
  {Laughlin}},\ }\bibfield  {title} {\enquote {\bibinfo {title} {Quantized hall
  conductivity in two dimensions},}\ }\href {\doibase 10.1103/PhysRevB.23.5632}
  {\bibfield  {journal} {\bibinfo  {journal} {Phys. Rev. B}\ }\textbf {\bibinfo
  {volume} {23}},\ \bibinfo {pages} {5632--5633} (\bibinfo {year}
  {1981})}\BibitemShut {NoStop}%
\bibitem [{\citenamefont {Sheng}\ \emph {et~al.}(2003)\citenamefont {Sheng},
  \citenamefont {Wan}, \citenamefont {Rezayi}, \citenamefont {Yang},
  \citenamefont {Bhatt},\ and\ \citenamefont {Haldane}}]{sheng2003}%
  \BibitemOpen
  \bibfield  {author} {\bibinfo {author} {\bibfnamefont {D.~N.}\ \bibnamefont
  {Sheng}}, \bibinfo {author} {\bibfnamefont {Xin}\ \bibnamefont {Wan}},
  \bibinfo {author} {\bibfnamefont {E.~H.}\ \bibnamefont {Rezayi}}, \bibinfo
  {author} {\bibfnamefont {Kun}\ \bibnamefont {Yang}}, \bibinfo {author}
  {\bibfnamefont {R.~N.}\ \bibnamefont {Bhatt}}, \ and\ \bibinfo {author}
  {\bibfnamefont {F.~D.~M.}\ \bibnamefont {Haldane}},\ }\bibfield  {title}
  {\enquote {\bibinfo {title} {Disorder-driven collapse of the mobility gap and
  transition to an insulator in the fractional quantum hall effect},}\ }\href
  {\doibase 10.1103/PhysRevLett.90.256802} {\bibfield  {journal} {\bibinfo
  {journal} {Phys. Rev. Lett.}\ }\textbf {\bibinfo {volume} {90}},\ \bibinfo
  {pages} {256802} (\bibinfo {year} {2003})}\BibitemShut {NoStop}%
\bibitem [{\citenamefont {Sheng}\ \emph {et~al.}(2006)\citenamefont {Sheng},
  \citenamefont {Weng}, \citenamefont {Sheng},\ and\ \citenamefont
  {Haldane}}]{sheng2006}%
  \BibitemOpen
  \bibfield  {author} {\bibinfo {author} {\bibfnamefont {D.~N.}\ \bibnamefont
  {Sheng}}, \bibinfo {author} {\bibfnamefont {Z.~Y.}\ \bibnamefont {Weng}},
  \bibinfo {author} {\bibfnamefont {L.}~\bibnamefont {Sheng}}, \ and\ \bibinfo
  {author} {\bibfnamefont {F.~D.~M.}\ \bibnamefont {Haldane}},\ }\bibfield
  {title} {\enquote {\bibinfo {title} {Quantum spin-hall effect and
  topologically invariant chern numbers},}\ }\href {\doibase
  10.1103/PhysRevLett.97.036808} {\bibfield  {journal} {\bibinfo  {journal}
  {Phys. Rev. Lett.}\ }\textbf {\bibinfo {volume} {97}},\ \bibinfo {pages}
  {036808} (\bibinfo {year} {2006})}\BibitemShut {NoStop}%
\end{thebibliography}%

\end{document}